\documentclass[aps]{revtex4}
\usepackage{amsfonts}
\usepackage{amsmath}
\usepackage{amssymb,epsf}

\begin{document}

\title{Critical behavior of charged black holes in Gauss-Bonnet gravity's
rainbow}
\author{Seyed Hossein Hendi$^{1,2}$\footnote{%
email address: hendi@shirazu.ac.ir}, Shahram
Panahiyan$^{1,3}$\footnote{ email address:
sh.panahiyan@gmail.com}, Behzad Eslam Panah$^{1}$\footnote{ email
address: behzad.eslampanah@gmail.com}, Mir Faizal$^{4}$\footnote{
email address: f2mir@uwaterloo.ca} and Mehrab
Momennia$^{1}$\footnote{ email address: momennia1988@gmail.com}}
\affiliation{$^1$ Physics Department and Biruni Observatory,
College of Sciences, Shiraz
University, Shiraz 71454, Iran\\
$^2$ Research Institute for Astronomy and Astrophysics of Maragha
(RIAAM), P.O. Box 55134-441, Maragha, Iran\\
$^3$ Physics Department, Shahid Beheshti University, Tehran 19839, Iran\\
$^4$ Department of Physics and Astronomy, University of
Lethbridge, Lethbridge, Alberta T1K 3M4, Canada}

\begin{abstract}
Following an earlier study regarding Gauss-Bonnet-Maxwell black
holes in the presence of gravity's rainbow [S. H. Hendi and M.
Faizal, Phys. Rev. D 92, 044027 (2015)], in this paper, we will
consider all constants as energy dependent ones. The geometrical
and thermodynamical properties of this generalization are studied
and the validation of the first law of thermodynamics is examined.
Next, through the use of proportionality between cosmological
constant and thermodynamical pressure, van der Waals-like behavior
of these black holes in extended phase space is investigated. An
interesting critical behavior for sets of rainbow functions in
this case is reported. Also, the critical behavior of uncharged
and charged solutions is analyzed and it is shown that the
generalization to a charged case puts an energy dependent
restriction on values of different parameters.
\end{abstract}

\maketitle

\section{Introduction}

Supergravity can be obtained as a low energy field theory
approximation of string theory \cite{lowe1,lowe}. The leading
order correction to this theory depends on the string used to
obtain the low energy effective field theory expansion. In the
heterotic string theory, the lowest corrections are described by a
Gauss-Bonnet (GB) term \cite{st12,12st,1s2t,s1t2}. Furthermore, it
is known from the renormalization group flow that the constants
depend on the scale at which a theory is probed \cite{renom}. So,
we expect that the different constants in general relativity and
GB gravity also depend on the scale at which these theories are
measured. The renormalization group flow for supergravity
solutions \cite{renor} and GB gravity \cite{regb} has been
analyzed. In fact, a renormalization group flow has been used for
measuring the flow of the cosmological constant \cite{cosmo} and
Newton constant \cite{newt}. Now, as the scale at which a theory
is measured depends on the energy of the probe, it is expected
that these constants will also depend on the energy. Therefore, in
this paper, we will use gravity's rainbow \cite{Magueijo} for
analyzing GB black holes with the energy dependent constants. It
may be noted that the initial motivation of gravity's rainbow came
by analyzing the target space metric of string theory. This was
done by regarding string theory as a two dimensional theory, and
considering the target space metric as a matrix of coupling
constants. Then using renormalization group flow, this matrix of
coupling constants would flow and depend on the scale at which
this theory is measured. This would in turn make them dependent on
the energy that is used for probing this theory \cite{Magueijo}.
Thus, the geometry would also be energy dependent.

Such a modification of a geometry at high energy scale can be viewed as a UV
completion. It may be noted that just like the Horava-Lifshitz gravity \cite%
{HoravaPRD,HoravaPRL}, the gravity's rainbow \cite{Magueijo}, has
also been viewed as a UV completion of general relativity. This is
because both of these approaches are based on a modification of
the usual energy-momentum dispersion relation in the UV limit.
Such a modification of the usual energy-momentum dispersion
relation in the UV limit occurs in a large number
of approaches to quantum gravity, such as the discrete spacetime \cite{Hooft}%
, models based on string field theory \cite{Samuel1}, spacetime foam \cite%
{Ellis}, spin-network in loop quantum gravity \cite{Gambini},
non-commutative geometry \cite{Carroll,FaizalMPLA} and ghost
condensation \cite{FaizalJPA}. It may be noted that even the
Greisen-Zatsepin-Kuzmin limit (GZK limit) suggests that the usual
energy-momentum relation could get modified in the UV limit
\cite{Greisen,Zatsepin}. The GZK limit can be used for analyzing
quantum gravitational effects as an upper limit on the energy of
cosmic ray. The Pierre Auger Collaboration and the High Resolution
Fly's Eye experiment have reconfirmed earlier results of the GZK
cutoff \cite{Abraham}. All these observations suggest that there
is a strong experimental motivation for a UV modification of the
usual energy-momentum dispersion relation. Motivated by the
Horava-Lifshitz gravity, the geometries occurring in the types IIA
\cite{Gregory} and IIB string theory \cite{Burda} have been also
modified in the UV limit. Different Lifshitz scaling for space and
time have also been used for analyzing certain aspects of the
AdS/CFT correspondence \cite{Gubser,Chen,Alishahiha,Kachru}. It
may be noted that for a suitable choice of the rainbow functions,
the
Horava-Lifshitz gravity can be related to the gravity's rainbow \cite%
{Garattini}. This is because both of these approaches are based on
a modification of general relativity in the UV limit such that the
general relativity is obtained in the IR limit. Basically,
Horava-Lifshitz gravity has been used to study the UV completion
of various interesting geometries. Motivated by the close relation
between the gravity's rainbow and Horava-Lifshitz gravity, the UV
completion of many black hole solutions has
been recently obtained by using the formalism of gravity's rainbow \cite%
{Ali341}

One of the most interesting effects of the UV completion of
geometries is the modification of the black hole thermodynamics at
the last stage of the evaporation of the black holes. As the
gravity's rainbow reduces to the general relativity in the IR
limit, the black hole thermodynamics in gravity's rainbow reduces
to the usual black hole thermodynamics for very large black holes.
However, as the black holes evaporate and reduce in size, the
black hole thermodynamics in gravity's rainbow showing deviation
from the usual black hole thermodynamics. This deviation becomes
significant at the
end stage of the evaporation of black hole in gravity's rainbow \cite%
{Ali104040}. It has been observed that the temperature of black holes
reaches a maximum value, and then it starts to reduce beyond this maximum
value. At a critical value the temperature of black holes becomes zero, and
the black holes do not radiate Hawking radiation at that stage. So, a black
hole remnant forms in gravity's rainbow. This formation of a black hole
remnant has important phenomenological consequences for the detection of
mini black holes at the LHC \cite{Ali295}. It is known that the usual
uncertainty principle has to be modified to the generalized uncertainty
principle to have an upper bound on the energy of a particle. As this
particle acts as a probe for the geometry of the black hole, it also fixes
the energy scale of the gravity's rainbow. Thus, this bound on the energy of
a particle emitted in the Hawking radiation can be used as energy scale in
the rainbow functions. This modifies the thermodynamics of black holes \cite%
{Ali104040,HPEMrainbow}. Such a modification in the black hole
thermodynamics has also been observed for black rings
\cite{Ali159}. The temperature of black rings also reaches a
maximum value and reduces beyond that value to form a remnant. It
has been argued that a remnant will form for all black objects
\cite{Ali341}. This has been explicitly demonstrated for Kerr
black holes, Kerr-Newman black holes in de Sitter space, charged
anti-de Sitter (AdS) black holes, higher dimensional Kerr-AdS
black holes and black saturn \cite{Ali341}. The geometric and
thermodynamic properties of the charged dilatonic black holes in
gravity's rainbow have also been investigated \cite{HFEP}.

The effect of gravity's rainbow on the thermodynamics of black
holes in GB gravity coupled to Maxwell's theory has been studied
\cite{HendiFaizalGB}. In this analysis, it was observed that even
though the thermodynamics of the black holes get modified in the
GB gravity's rainbow, the first law of thermodynamics still holds
for this modified thermodynamics. However, in this analysis,
different constants were not made to depend on the energy of the
probe. As we expect the constant to depend on the energy of the
probe, in this paper, we will generalize this analysis and make
the constants energy dependant. We will also investigate the
critical behavior of black hole solutions in this theory. It may
be noted that the pressure-volume (PV) criticality has been
studied for black holes in GB gravity using extended phase space
thermodynamics \cite{Mo,Cao,Wei,Zou}. In the extended phase space
thermodynamics, the cosmological constant is viewed as a
thermodynamic pressure, and so it is possible to define a
thermodynamic volume conjugate to this thermodynamic pressure
\cite{Kubiznak,OurPV1,OurPV2,OurPV3,OurPV4}. It has been observed
that along with the usual black hole phase transition, a new
phenomenon of reentrant phase transitions occurs for rotating AdS
black holes in extended phase space \cite{Altamirano}. In these
reentrant phase transitions a monotonic variation of the
temperature yields two phase transitions, and this situation is
similar to that which is seen in multi-component liquids. The PV
criticality has also been studied in quasi-topological gravity
\cite{pv}. In this paper, we study the effect of the UV completion
of general relativity on the PV criticality in extended phase
space. Also, we analyze the effects of rainbow functions on PV
criticality in GB gravity's rainbow. In addition, we analyze the
effect of energy dependent constants on thermodynamic properties
of black hole solutions. We also study the effects of such
modification on the critical values and van der Waals like
behavior of the system.

\section{Probe energy dependent constants: exact solutions}

As it was mentioned before, we expect that all the constants
depend on the energy of the probe. Such a dependency of the
constants on the energy of the probe can be explicitly analyzed
using gravity's rainbow. This is an advantage of using the
gravity's rainbow. Now, following earlier work on
Einstein-GB-Maxwell black holes in gravity's rainbow, we will
generalize solutions to the case where different constants are
dependent on the energy of the probe. Our first step is analyzing
the effect of this energy dependency on the field equations.

The $d$-dimensional action of GB-Maxwell gravity with negative cosmological
constant can be written as
\begin{equation}
\mathcal{I}=-\frac{1}{2\kappa }\int d^{d}x\sqrt{-g}[\alpha _{0}\mathcal{L}%
_{0}+\alpha _{1}\mathcal{L}_{1}+\alpha _{2}\mathcal{L}_{2}+\kappa \mathcal{L}%
(\mathcal{F})],  \label{Action}
\end{equation}%
where $\kappa =8\pi G(E)$ and we set $\alpha _{0}=\alpha _{1}=1$
and $\alpha _{2}=\alpha (E)$ in which the last one is the
so-called GB coefficient. In this paper, we will only consider
positive values of GB coefficient. In addition,
$\mathcal{L}_{i}$'s are the first three terms of Lovelock
Lagrangian which are corresponding to the cosmological constant,
Einstein and GB Lagrangian, with the following explicit forms
\begin{eqnarray*}
\mathcal{L}_{0} &=&-2\Lambda (E), \\
\mathcal{L}_{1} &=&\mathcal{R}, \\
\mathcal{L}_{2} &=&R_{\mu \nu \gamma \delta }R^{\mu \nu \gamma \delta
}-4R_{\mu \nu }R^{\mu \nu }+\mathcal{R}^{2}.
\end{eqnarray*}

The $\mathcal{L}(\mathcal{F})$ is the Lagrangian of electrodynamics in which
we choose linear Maxwell Lagrangian, $\mathcal{L}(\mathcal{F})=-\mathcal{F}$, where $%
\mathcal{F}=F_{\mu \nu }F^{\mu \nu }$ is the Maxwell invariant.

Variation of the action (\ref{Action}) with respect to the metric tensor $%
g_{\mu \nu }$ and the Faraday tensor $F_{\mu \nu }$, leads to the following
field equations%
\begin{equation}
G_{\mu \nu }^{0}+G_{\mu \nu }^{1}+\alpha _{2}(E) G_{\mu \nu }^{2}=8\pi G(E)
T_{\mu \nu },  \label{Field equation}
\end{equation}
\begin{equation}
\nabla _{\mu }F^{\mu \nu }=0,  \label{Maxwell equation}
\end{equation}%
where
\begin{eqnarray}
T_{\mu \nu } &=&2F_{\mu \lambda }F_{\nu }^{\lambda }-\frac{1}{2}g_{\mu \nu }%
\mathcal{F}, \\
G_{\mu \nu }^{0} &=&-\frac{1}{2}g_{\mu \nu }\mathcal{L}_{0},  \label{G0} \\
G_{\mu \nu }^{1} &=&R_{\mu \nu }-\frac{1}{2}g_{\mu \nu }\mathcal{L}_{1},
\label{G1} \\
G_{\mu \nu }^{2} &=&-2\left( 2R^{\rho \sigma }R_{\mu \rho \nu \sigma
}-R_{\mu }^{\rho \sigma \lambda }R_{\nu \rho \sigma \lambda }-RR_{\mu \nu
}+2R_{\mu \lambda }R_{\nu }^{\lambda }\right) -\frac{1}{2}g_{\mu \nu }%
\mathcal{L}_{2}.  \label{G2}
\end{eqnarray}

Following Ref. \cite{HendiFaizalGB}, one finds the spherical symmetric
metric governing the gravity's rainbow which has dependency on rainbow
functions in following form
\begin{equation}
d\tau =-ds^{2}=\frac{\Psi \left( r\right) }{f(E)
^{2}}dt^{2}-\frac{1}{g(E) ^{2}}\left[ \frac{dr^{2}}{\Psi \left(
r\right) }+r^{2}\left( d\theta
_{1}^{2}+\sum\limits_{i=2}^{d-2}\prod\limits_{j=1}^{i-1}\sin
^{2}\theta _{j}d\theta _{i}^{2}\right) \right] ~.  \label{Metric}
\end{equation}

Using Eqs. (\ref{Field equation}) - (\ref{G2}) with metric (\ref{Metric}),
we can find following electromagnetic field tensor and metric function
\begin{eqnarray}
F_{tr} &=&\frac{q(E)}{r^{d_{2}}}, \\
\Psi \left( r\right) &=&1+\frac{r^{2}}{2\alpha ^{\prime }(E)g^{2}(E)}\left(
1-\sqrt{\Theta \left( r\right) }\right) ,  \label{metric function} \\
\Theta \left( r\right) &=&1+\frac{8\alpha ^{\prime }(E)}{d_{1}d_{2}}\left(
\Lambda (E)+\frac{d_{1}d_{2}m(E)}{2r^{d_{1}}}-\frac{8d_{1}d_{3}\pi
G(E)q^{2}(E)g^{2}(E)f^{2}(E)}{r^{2d_{2}}}\right) ,
\end{eqnarray}%
where $d_{i}=d-i$ . Also, $q(E)$ and $m(E)$ are integration constants which
are, respectively, related to the total electric charge and total mass of
the solutions, and $\alpha ^{\prime }(E)=d_{3}d_{4}\alpha (E)$.

For the simplicity and in order to find the contribution of each
parameters, we consider following notations
\begin{equation}
q(E)=h_{1}^{2}(E)q,\;\;\;\Lambda (E)=h_{2}^{2}(E)\Lambda
,\;\;\;G(E)=h_{3}^{2}(E)G,\;\;\;\alpha ^{\prime }(E)=h_{4}^{2}(E)\alpha
^{\prime },\;\;\;m(E)=h_{5}^{2}(E)m,
\end{equation}%
and since we are working in natural units, we set $8\pi G=1$.

Here, we would like to make some remarks regarding the properties of the
solutions. Calculations show that there is a curvature singularity at $r=0$,
which can be covered with an event horizon. Hence, it will be interesting to
see the effects of the gravity's rainbow on the singularity and asymptotical
behavior of the solutions. In order to study these effects, we use series
expansion of the Kretschmann scalar for small and large values of radial
coordinate. By doing so, one can find following relations
\begin{equation}
{\lim_{r\longrightarrow 0}}R_{\alpha \beta \gamma \delta }R^{\alpha \beta
\gamma \delta }=\frac{-4d_{3}(d^{2}-7d+13)g(E)
^{d_{3}}f(E)^{2}q^{2}h_{1}^{4}(E)}{\alpha ^{\prime }h_{4}^{2}(E)r^{2d_{2}}},
\label{RR0}
\end{equation}%
\begin{equation}
{\lim_{r\longrightarrow \infty }}R_{\alpha \beta \gamma \delta }R^{\alpha
\beta \gamma \delta }=g(E) ^{d_{1}}\left[ \frac{8d_{-1}\Lambda _{eff}^{2}}{%
d_{1}^{2}d_{2}}-\frac{4\Lambda _{eff}}{d_{1}d_{2}}\right] ,  \label{RRinf}
\end{equation}%
where%
\begin{equation}
\Lambda _{eff}=\frac{d_{1}d_{2}}{4\alpha ^{\prime }h_{4}^{2}(E)g(E)^{2}}%
\left[ \sqrt{1+\frac{8\alpha ^{\prime }h_{4}^{2}(E)\Lambda h_{2}^{2}(E) }{%
d_{1}d_{2}}}-1\right] .  \label{Leff}
\end{equation}

Interestingly, the obtained relations indicate that the rainbow
functions affect the strength of singularity as well as
asymptotical behavior of the solutions. In other words, the
asymptotical behavior of the solutions is AdS with an effective
cosmological constant, $\Lambda _{eff}$. It is clear that the GB
parameter and rainbow function can modify $\Lambda _{eff}$. It
means that $\Lambda _{eff}$ reduces to $\Lambda (E)$ for $\alpha
(E)\rightarrow 0$ and $g(E)\rightarrow 1$, simultaneously.

\section{Thermodynamical quantities}

Now, we are in a position to study thermodynamical quantities of the
solutions. Using the concept of the surface gravity, it is a matter of
calculation to obtain temperature as
\begin{equation}
T=\frac{\left[ d_{2}g^{2}(E)\left( d_{3}r_{+}^{2}+\alpha ^{\prime }
d_{5}h_{4}^{2}(E) g^{2}(E) \right) -2\Lambda h_{2}^{2}(E) r_{+}^{4}\right]
r_{+}^{2d}-2q^{2}d_{3}^{2}h_{1}^{4}(E) h_{3}^{2}(E) g^{2}(E) f^{2}(E)
r_{+}^{8}}{4\pi d_{2}\left( r_{+}^{2}+2\alpha ^{\prime } h_{4}^{2}(E)
g^{2}(E) \right) g(E)f(E)r_{+}^{2d_{-1/2}}}.  \label{TNEW}
\end{equation}

One of the conditions for having physical solutions is imposed by positive
temperature. Since we have considered positive values of GB parameter, the
denominator of the temperature is always positive and only numerator has
contribution to negativity of the solutions. For AdS black holes, only
charge term in numerator of the temperature contribute to negativity of it.
Taking closer look at the numerator, one can see that energy dependency of
constants plays a crucial role in domination of different terms. Therefore,
the conditions for having physical solutions (positive temperature) is
highly sensitive to energy variation of different constants.

Since we are working in the context of higher derivative gravity,
it is not allowed to use the area law for calculating entropy. We
use the Wald formula with the following result
\begin{equation}
S=\frac{r_{+}^{d_{2}}\left( 1+\frac{2\alpha ^{\prime } d_{2}h_{4}^{2}(E)
g^{2}(E) }{d_{4}r_{+}^{2}}\right) }{4g^{d_{2}}(E) }.  \label{SNEW}
\end{equation}

In addition, the total electric charge of the solutions is obtained through
the use of Gauss's law as
\begin{equation}
Q=\frac{qd_{3}h_{1}^{2}(E) h_{3}^{2}(E) f(E) }{4\pi g^{d_{3}}(E)}.
\label{QNEW}
\end{equation}

For the electric potential, we have
\begin{equation}
U=A_{\mu }\chi ^{\mu }\left\vert _{r\rightarrow \infty }\right. -A_{\mu
}\chi ^{\mu }\left\vert _{r\rightarrow r_{+}}\right.=\frac{h_{1}^{2}(E) q}{%
r_{+}^{d_{3}}}.  \label{UNEW}
\end{equation}

The total mass of the black holes could be obtained through
counter-term method or the Arnowitt-Deser-Misner approach with the
following unique form
\begin{equation}
M=\frac{1}{16\pi }\frac{h_{5}^{2}(E)d_{2}m}{f(E)g^{d_{1}}(E)},  \label{M2}
\end{equation}%
where by evaluating metric function on outer horizon, one can find
total finite mass as
\begin{equation}
M=\frac{\left[ \frac{1}{2}d_{1}d_{2}g^{2}(E)\left( r_{+}^{2}+\alpha ^{\prime
}h_{4}^{2}(E)g^{2}(E)\right) -\Lambda r_{+}^{4}h_{2}^{2}(E)\right]
r_{+}^{d_{5}}+q^{2}d_{1}d_{3}h_{1}^{4}(E)h_{3}^{2}(E)g^{2}(E)f^{2}(E)r_{+}^{-2d_{2}}%
}{8\pi d_{1}f(E)g^{d_{1}}(E)}.  \label{MNEW}
\end{equation}

Before we proceed, it is worthwhile to mention a few remarks about
conserved and thermodynamical quantities. The modification of GB
gravity when the constants are not energy dependent has already
been analyzed \cite{HendiFaizalGB}. Thermodynamical quantities of
the solutions in \cite{HendiFaizalGB} differ from those obtained
here which highlights the effects of dependency of constants on
the probe energy. Especially, the electric potential was shown to
be independent of energy function in Ref. \cite{HendiFaizalGB},
while here, it depends on the energy of the probe. Strictly
speaking, these two types of black holes are phenomenologically
different. Recently, it was pointed out that most of constants in
physics are not actually constant. In fact, a measurement of their
expectation values leads to the result that they should be varying
parameters. Here, we have taken such consideration into account
and shown that thermodynamically speaking, black holes will be
modified in such consideration.

Now, we are in a position to check the validation of the first law
of black hole's thermodynamics. Using total mass of these black
holes (\ref{MNEW}) with the obtained entropy (\ref{SNEW}) and
electric charge (\ref{QNEW}) as extensive parameters, one can
define following intensive parameters

\begin{equation}
T=\left( \frac{\partial M}{\partial r_{+}}\right) _{q}\ \left( \frac{%
\partial r_{+}}{\partial S}\right) _{q}\ \ \ \ \ \ and\ \ \ \ \ \ \ \ \
U=\left( \frac{\partial M}{\partial q}\right) _{r_{+}},  \label{TU}
\end{equation}%
where by evaluating these equations, one can confirm that the
obtained temperature and electric potential in Eq. (\ref{TU})
coincide with those extracted from Eqs. (\ref{TNEW}) and
(\ref{UNEW}). Therefore, despite the modifications in
thermodynamical quantities by consideration of energy dependant
constants, the first law of thermodynamics is valid for these
black holes.

\section{Probe energy dependent constants: \newline
critical quantities and van der Waals like behavior \label{PV}}

The extended phase space expression comes from consideration of
the cosmological constant not as a fixed quantity but a
thermodynamical variable which is known as pressure. Although
thermodynamical pressure is generally proportional to the negative
cosmological constant with a proportionality constant
$\frac{-1}{8\pi }$, there are some cases in which we should modify
it \cite{HendiArmanBD,HFEP}. In this paper, it is observed that
the metric function and asymptotical behavior of the system have
been modified due to the existence of rainbow functions.
Therefore, it is necessary to check the possible effects of
rainbow functions on thermodynamical pressure. To do so, we
evaluate the energy-momentum tensor. It is straightforward to
obtain the following relations for different components of
energy-momentum tensor
\begin{equation}
T_{t}^{t}=T_{r}^{r}=-\frac{f^{2}(E) g^{2}(E) F_{tr}^{2}}{8\pi }-\frac{%
\Lambda (E)}{8\pi },
\end{equation}%
\begin{equation}
T_{\theta _{i}}^{\theta _{i}}=\frac{f^{2}(E) g^{2}(E) F_{tr}^{2}}{8\pi }-%
\frac{\Lambda (E)}{8\pi }.
\end{equation}

In these relations, the first term is due to the existence of
electromagnetic field (which is coupled with rainbow functions).
Surprisingly, the $\Lambda (E)$ term is not coupled with any
rainbow functions of the metric. In other words, although rainbow
functions of the metric modify $\Lambda (E)$ term in the metric
function and asymptotical behavior of the solutions, they do not
affect thermodynamical pressure which is related to the
cosmological constant. Therefore, in studying the critical
behavior of the system through the analogy between $\Lambda $ and
$P$, one can use following relation
\begin{equation}
P=-\frac{\Lambda (E)}{8\pi }=-\frac{h_{2}^{2}(E) \Lambda}{8\pi }.
\label{PNEW}
\end{equation}

Thermodynamically speaking, the conjugating thermodynamical
variable corresponding to the pressure is thermodynamical volume
which in the context of enthalpy is calculated by
\begin{equation}
V=\left( \frac{\partial H}{\partial P}\right) _{S,Q}.  \label{V}
\end{equation}

It is worthwhile to mention that in order to have a well-defined vacuum
solution with $m=q=0$, the pressure $P$ has to satisfy the following
constraint \cite{Cao,MaximallPressure2,MaximallPressure3}
\begin{equation}
0\leq \frac{64\pi \alpha ^{\prime }h_{4}^{2}(E)P}{d_{1}d_{2}}\leq 1,
\end{equation}%
in which it puts a restriction on the pressure as maximal pressure
\begin{equation}
P\leq P_{\max }=\frac{d_{1}d_{2}}{64\pi \alpha ^{\prime }h_{4}^{2}(E)},
\label{maximall}
\end{equation}%
which indicates that only for sufficiently small pressures, the solution (%
\ref{metric function}) has an asymptotic AdS region.

Now, remembering that we are working in extended phase space, the total mass
of the black holes will not play the role of internal energy. Instead, it is
interpreted as enthalpy. With this consideration, one can find the Gibbs
free energy as
\begin{equation}
G=H-TS=M-TS.  \label{G}
\end{equation}

As for the volume of these black holes, by using Eqs. (\ref{MNEW}), (\ref%
{PNEW}) and (\ref{V}), we obtain
\begin{equation}
V=\frac{r_{+}^{d_{1}}}{d_{1}g^{d_{1}}(E) f(E) },
\end{equation}
which is modified in the presence of the rainbow functions. In
other words, in the presence of gravity's rainbow, thermodynamical
volume of the black holes is a function of the rainbow functions,
and therefore, the final form of these black holes is determined
by the model of rainbow functions under consideration. On the
other hand, even by consideration of the dependency of different
constants on the probe energy, the volume of these black holes is
same as that of probe energy independent constants. This behavior
is expected, since the abstract form of metric (\ref{Metric}) is
free of any constant.

The equation of state and Gibbs free energy of these solutions could be
found by using Eqs. (\ref{TNEW}), (\ref{PNEW}) and (\ref{G}), which result
into
\begin{eqnarray}
P &=&\frac{d_{2}\left[ r_{+}^{2}+2\alpha ^{\prime }h_{4}^{2}(E)g^{2}(E)%
\right] g(E)f(E)}{4r_{+}^{3}}T+\frac{%
d_{3}^{2}h_{1}^{4}(E)h_{3}^{2}(E)g^{2}(E)f^{2}(E)}{8\pi r_{+}^{2d_{2}}}q^{2}
\notag \\
&&-\frac{d_{2}\left[ d_{3}r_{+}^{2}+\alpha ^{\prime
}d_{5}h_{4}^{2}(E)g^{2}(E)\right] g^{2}(E)}{16\pi r_{+}^{4}},  \label{PPNEW}
\end{eqnarray}%
and
\begin{eqnarray}
G &=&\frac{1}{d_{4}g(E)f(E)\left( r_{+}^{2}+2\alpha ^{\prime
}h_{4}^{2}(E)g^{2}(E)\right) r_{+}^{4}}\left\{ \frac{d_{4}r_{+}^{d_{-3}}}{%
16\pi g^{d_{4}}(E)}+\frac{\alpha ^{\prime }}{8\pi }\left( \frac{%
d_{8}r_{+}^{d_{-1}}}{2g^{d_{6}}(E)}+\frac{\alpha ^{\prime
}d_{2}h_{4}^{2}(E)r_{+}^{d_{1}}}{g^{d_{8}}(E)}\right) h_{4}^{2}(E)\right.
\notag \\
&&\left. -\frac{6P}{d_{1}}\left( \frac{d_{4}r_{+}^{d_{-5}}}{%
6d_{2}g^{d_{2}}(E)}+\frac{\alpha ^{\prime }h_{4}^{2}(E)r_{+}^{d_{-3}}}{%
g^{d_{4}}(E)}\right) +\frac{d_{3}q^{2}h_{1}^{4}(E)h_{3}^{2}(E)f^{2}(E)}{2\pi
d_{2}}\left( \frac{d_{4}d_{5/2}}{2g^{d_{4}}(E)r_{+}^{d_{9}}}+\frac{\alpha
^{\prime }d_{2}d_{7/2}h_{4}^{2}(E)}{g^{d_{6}}(E)r_{+}^{d_{7}}}\right)
\right\} .  \label{GNEW}
\end{eqnarray}

There are several approaches toward calculating critical values. In this
paper, we will employ the properties of the inflection points in $P-r_{+}$
diagrams. In this method, one can follow the relations for calculating
critical volume which in case of these black holes, it will be proportional
to the horizon radius%
\begin{equation}
\left( \frac{\partial P}{\partial r_{+}}\right) _{T}=\left( \frac{\partial
^{2}P}{\partial r_{+}^{2}}\right) _{T}=0,  \label{infel}
\end{equation}%
where by using Eqs. (\ref{PPNEW}) and (\ref{infel}), one obtains
\begin{equation}
\frac{4qd_{3}^{2}h_{1}^{4}(E) h_{3}^{2}(E) g(E)f^{2}(E) \left[
d_{5/2}+6\alpha ^{\prime } h_{4}^{2}(E) d_{7/2}g^{2}(E) r_{+}^{2}\right] }{%
r_{+}^{2d_{5}}}-d_{3}g(E)r_{+}^{4}+12\alpha ^{\prime } h_{4}^{2}(E) g^{3}(E) %
\left[ r_{+}^{2}-\alpha ^{\prime } d_{5}h_{4}^{2}(E) g^{2}(E) \right] =0.
\end{equation}

As a special case, we consider $5$-dimensional solutions in the absence of
electric charge to obtain critical horizon radius, temperature and pressure.
So, we can write
\begin{equation}
r_{c}=\sqrt{6\alpha ^{\prime }}h_{4}(E) g(E) ,\;\;\;\; \& \;\;\;\;T_{c}=%
\frac{1}{2\pi h_{4}(E) f(E) \sqrt{6\alpha ^{\prime }}}, \;\;\;\;\; \&
\;\;\;\;\;P_{c}=\frac{1}{48h_{4}^{2}(E) \alpha ^{\prime }\pi},
\end{equation}
which lead to the following ratio
\begin{equation}
\frac{P_{c}r_{c}}{T_{c}}=\frac{f(E) g(E) }{4}.  \label{PVCNEW}
\end{equation}

It is notable that in the absence of rainbow functions ($f(E)=g(E)=1$), this
ratio reduces to $1/4$, and therefore, Eq. (\ref{PVCNEW}) indicates that
consideration of the gravity's rainbow can modify this near universal ratio.
Interestingly, for neutral solutions ($q=0$), the critical pressure is
independent of metric's rainbow functions while the critical horizon radius
and the critical temperature are functions of one of rainbow functions.

Next, we are going to consider the charged GB black holes in the
presence of gravity's rainbow in $5-$dimensions. The critical
horizon radius in this case is given as
\begin{equation}
r_{c-q}=\frac{\sqrt{6\mathcal{B}^{1/3}\left[ \mathcal{B}^{2/3}+3\alpha
^{\prime }h_{4}^{2}(E)g^{2}(E)\mathcal{B}%
^{1/3}+15q^{2}h_{1}^{4}(E)h_{3}^{2}(E)f^{2}(E)+9\alpha ^{\prime
2}h_{4}^{4}(E)g^{4}(E)\right] }}{3\mathcal{B}^{1/3}}.  \label{RCNEW}
\end{equation}%
where%
\begin{eqnarray*}
\mathcal{B} &\mathcal{=}&189\alpha ^{\prime
}q^{2}h_{1}^{4}(E)h_{3}^{2}(E)h_{4}^{2}(E)g^{2}(E)f^{2}(E)+27\alpha ^{\prime
3}h_{4}^{6}(E)g^{6}(E)+3qh_{1}^{2}(E)h_{3}(E)f(E)\times \\
&& \\
&&\sqrt{729\alpha ^{\prime
4}h_{4}^{8}(E)g^{8}(E)-375q^{4}h_{1}^{8}(E)h_{3}^{4}(E)f^{4}(E)+3294\alpha
^{\prime 2}q^{2}h_{1}^{4}(E)h_{3}^{2}(E)h_{4}^{4}(E)g^{4}(E)f^{2}(E)}.
\end{eqnarray*}

It is worthwhile to emphasize that, in this paper, we consider the
positive values of GB parameter ($\alpha^{\prime}>0$). The
negative value of $\alpha^{\prime}$ enforces other set of
conditions for having positive critical parameters. Furthermore,
$T_{c-q}$ will be given by
\begin{equation}
T_{c-q}=\frac{g(E)\left[ r_{c-q}^{4}-4q^{2}h_{1}^{4}(E)h_{3}^{2}(E)f^{2}(E)%
\right] }{\pi \left( r_{c-q}^{2}+6\alpha ^{\prime
}h_{4}^{2}(E)g^{2}(E)\right) f(E)r_{c-q}^{3}},  \label{TCNEW}
\end{equation}%
and $P_{c-q}$ is
\begin{equation}
P_{c-q}=\frac{g^{2}(E)\left\{
3r_{c-q}^{6}-4q^{2}h_{1}^{4}(E)h_{3}^{2}(E)f^{2}(E)\left[ 5r_{c-q}^{2}+6%
\alpha ^{\prime }h_{4}^{2}(E)g^{2}(E)\right] -6\alpha ^{\prime
}h_{4}^{2}(E)g^{2}(E)r_{c-q}^{4}\right\} }{8\pi \left( r_{c-q}^{2}+6\alpha
^{\prime }h_{4}^{2}(E)g^{2}(E)\right) r_{c-q}^{6}}.  \label{PCNEW}
\end{equation}

The obtained relation for $T_{c-q}$ imposes a restriction for
having positive critical temperature. Since denominator of
$T_{c-q}$ is positive, the restriction for having positive
$T_{c-q}$ comes from the numerator of this relation with the
following explicit form
\begin{equation}
r_{c-q}^{4}-4q^{2}h_{1}^{4}(E)h_{3}^{2}(E)f^{2}(E)>0.
\end{equation}

It is evident that the positivity of the critical temperature
depends on the energy variations of constants, the electric charge
and energy function of the metric. As for the critical pressure,
similarly, only its numerator may yield negative values. In other
words, the numerator of the critical pressure imposes following
condition for having positive values of $P_{c-q}$
\begin{equation}
3r_{c-q}^{6}-4q^{2}h_{1}^{4}(E)h_{3}^{2}(E)f^{2}(E)\left[ 5r_{c-q}^{2}+6%
\alpha ^{\prime }h_{4}^{2}(E)g^{2}(E)\right] -6\alpha ^{\prime
}h_{4}^{2}(E)g^{2}(E)r_{c-q}^{4}>0.
\end{equation}

Here, the restriction is rooted in two parts of the action; one is
related to the generalization of GB gravity and the other one is
related to the electrodynamic part. In other words, by cancelling
the contributions of GB gravity and electromagnetic field, the
pressure will always be positive.

Now, we are in a position to calculate $\frac{P_{c-q}r_{c-q}}{T_{c-q}}$.
Using Eqs. (\ref{RCNEW}), (\ref{TCNEW}) and (\ref{PCNEW}), we obtain
\begin{equation}
\frac{P_{c-q}r_{c-q}}{T_{c-q}}=\frac{g(E)f(E)\left\{
3r_{c-q}^{6}-4q^{2}h_{1}^{4}(E)h_{3}^{2}(E)f^{2}(E)\left[ 5r_{c-q}^{2}+6%
\alpha ^{\prime }h_{4}^{2}(E)g^{2}(E)\right] -6\alpha ^{\prime
}h_{4}^{2}(E)g^{2}(E)r_{c-q}^{4}\right\} }{8\left(
r_{c-q}^{4}-4q^{2}h_{1}^{4}(E)h_{3}^{2}(E)f^{2}(E)\right) r_{c-q}^{2}}.
\label{PVTNEWW}
\end{equation}

First of all, contrary to the absence of charge, in this case, all critical
values are functions of both of the rainbow functions of metric. This
emphasizes the fact that thermodynamical structure of the charged black
holes in gravity's rainbow is completely different from the neutral ones.

The critical horizon radius must be real valued and positive. Therefore,
there is a restriction
\begin{equation*}
729\alpha ^{\prime 4}h_{4}^{8}(E) g^{8}(E) -375q^{4}h_{1}^{8}(E)
h_{3}^{4}(E) f^{4}(E) +3294\alpha ^{\prime 2}q^{2}h_{1}^{4}(E) h_{3}^{2}(E)
h_{4}^{4}(E) g^{4}(E) f^{2}(E) >0.
\end{equation*}

We should note that this restriction is due to the existence of charge. In
other words, generalization from neutral to charged solutions, put
restrictions on values that rainbow functions and GB parameter can acquire.

%%%%%%%%%%%%%%%%%%%%%%%%%%%%%%%%%%%%%%%%%%%%%%%%%%%%%%%%%%%%%%%
\begin{figure}[tbp]
$%
\begin{array}{ccc}
\epsfxsize=5cm \epsffile{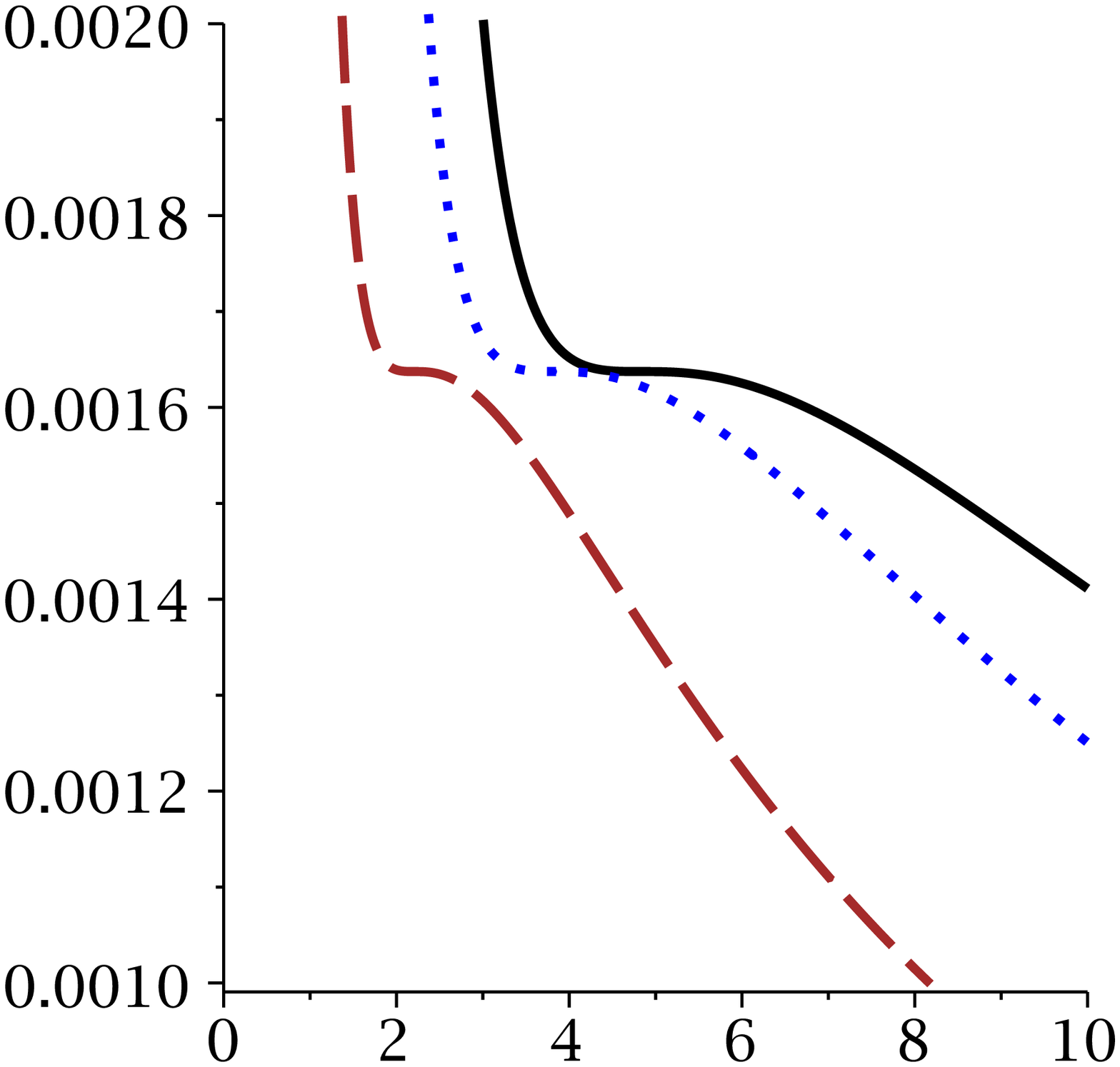} & \epsfxsize=5cm %
\epsffile{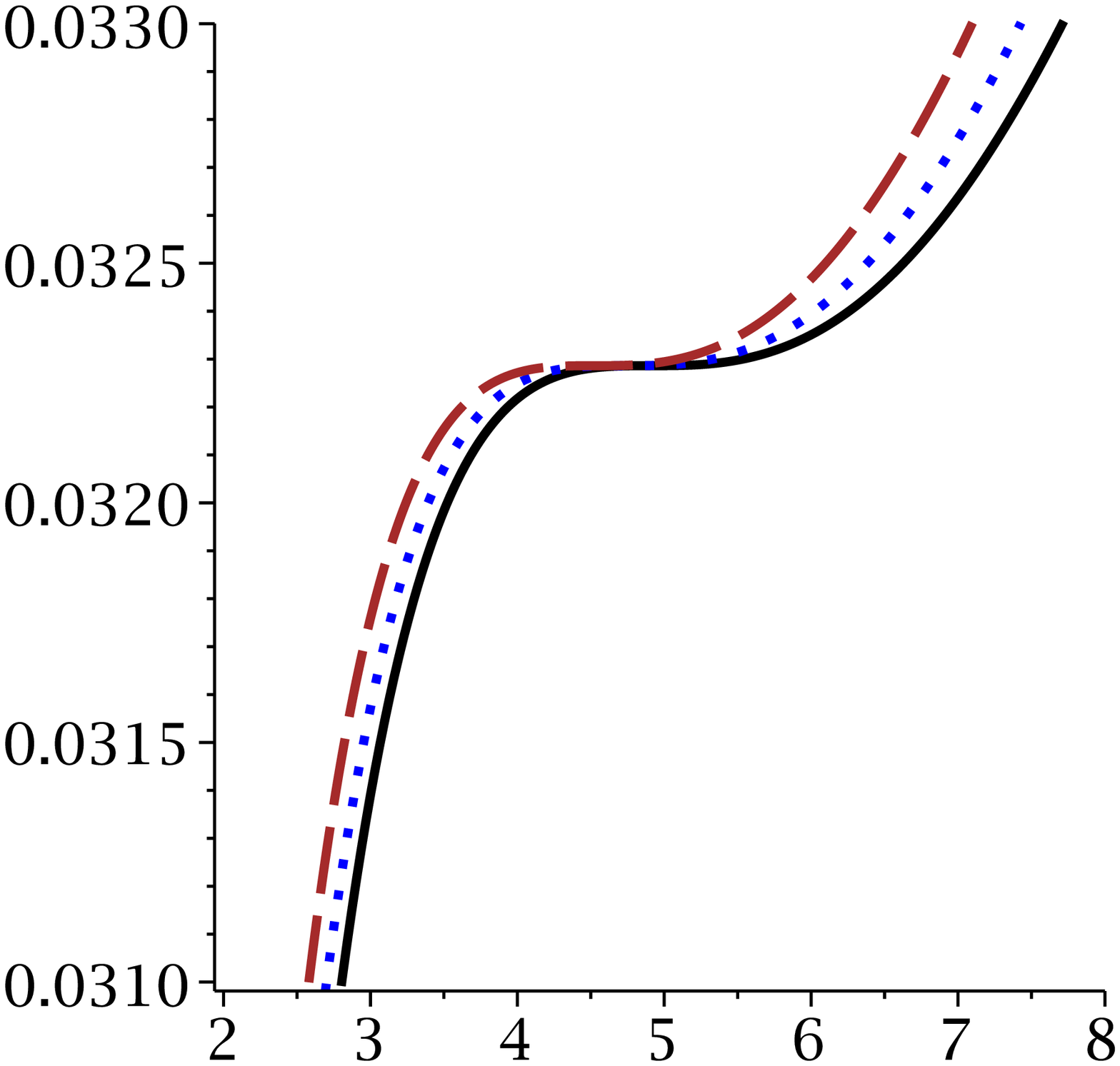} & \epsfxsize=5cm \epsffile{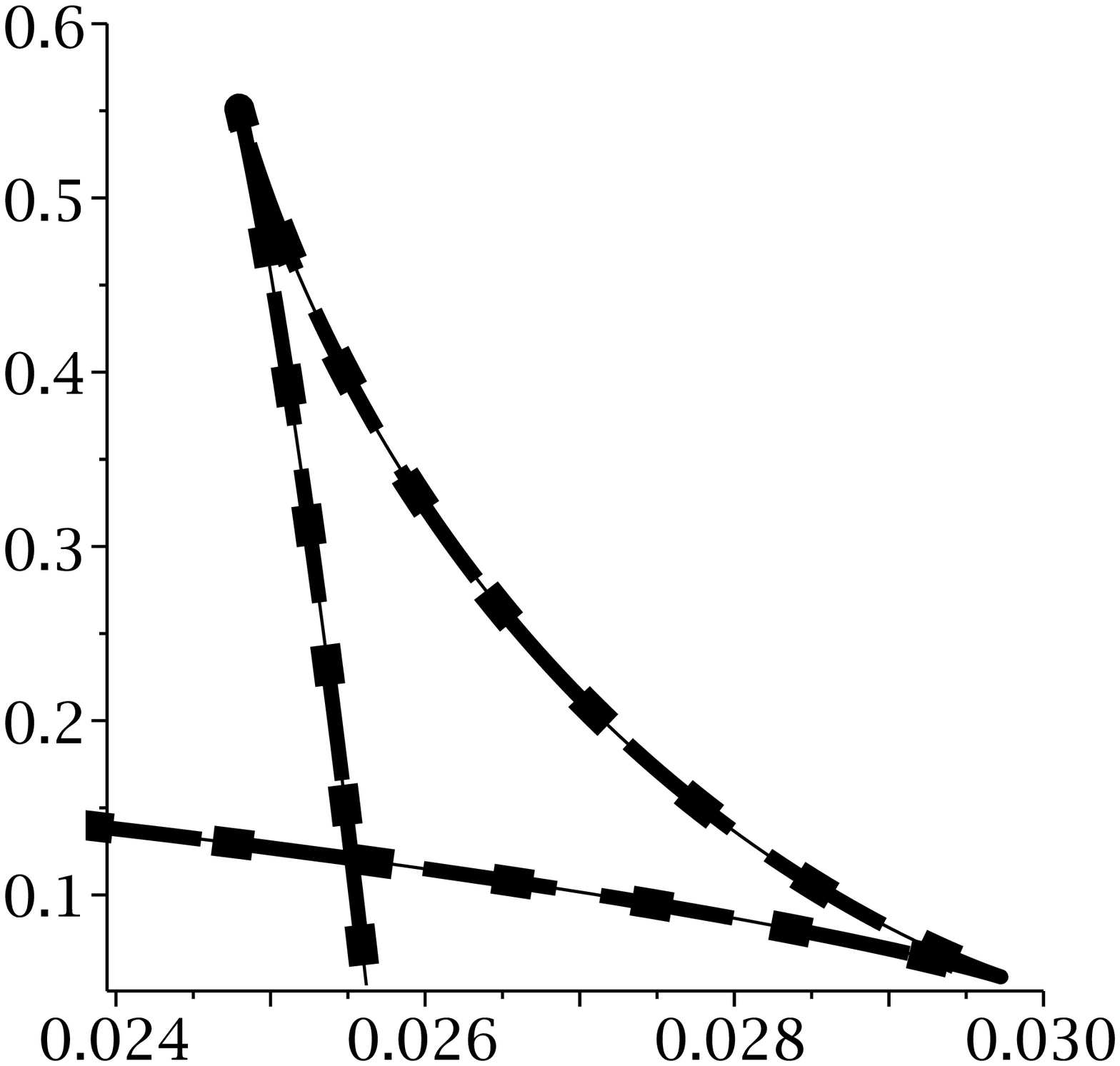}%
\end{array}
$%
\caption{ $P-r_{+}$ (left), $T-r_{+}$ (middle) and $G-T$ (right) diagrams
for $\protect\alpha ^{\prime }=5$, $q=0$ and $d=5$. \newline
$g\left(E/E_{p}\right) =\protect\sqrt{1-\protect\eta \left( E/E_{p}\right)
^{n}}$, $f\left( E/E_{p}\right) =1$, $h_{i}(E)=0.9$, $E=1$, $E_{p}=5$, $n=2$%
, $\protect\eta=1 $ (continuous line), $\protect\eta=10$ (dotted line) and $%
\protect\eta=20$ (dashed line). \newline
$P-r_{+}$ diagram for $T=T_{c}$, $T-r_{+}$ diagram for $P=P_{c}$ and $G-T$
diagram for $P=0.5P_{c}$. }
\label{Fig1}
\end{figure}
%%%%%%%%%%%%%%%%%%%%%%%%%%%%%%%%%%%%%%%%%%%%%%%%%%%%%%%%%%%%%%%
%%%%%%%%%%%%%%%%%%%%%%%%%%%%%%%%%%%%%%%%%%%%%%%%%%%%%%%%%%%%%%%
\begin{figure}[tbp]
$%
\begin{array}{ccc}
\epsfxsize=5cm \epsffile{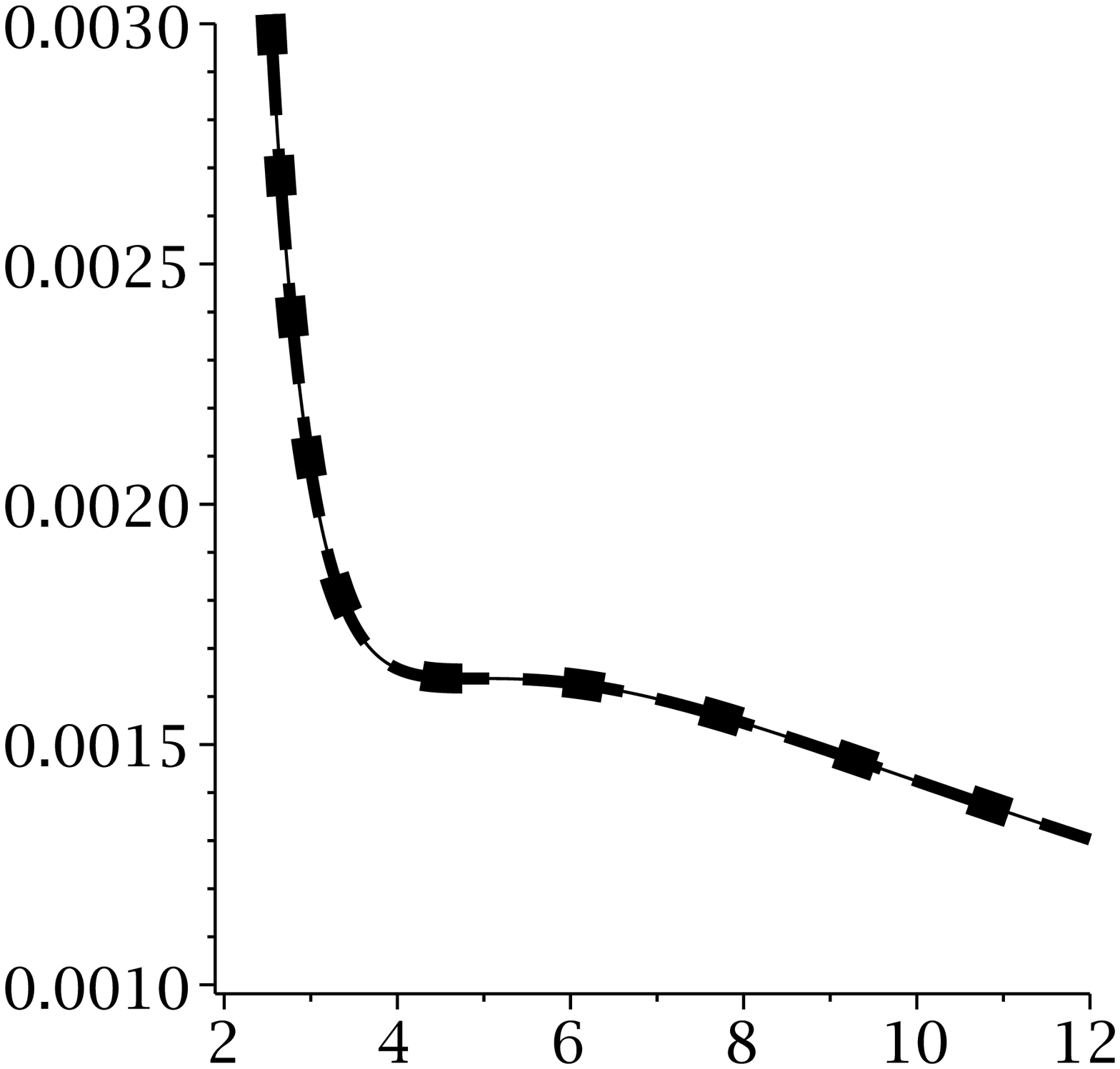} & \epsfxsize=5cm %
\epsffile{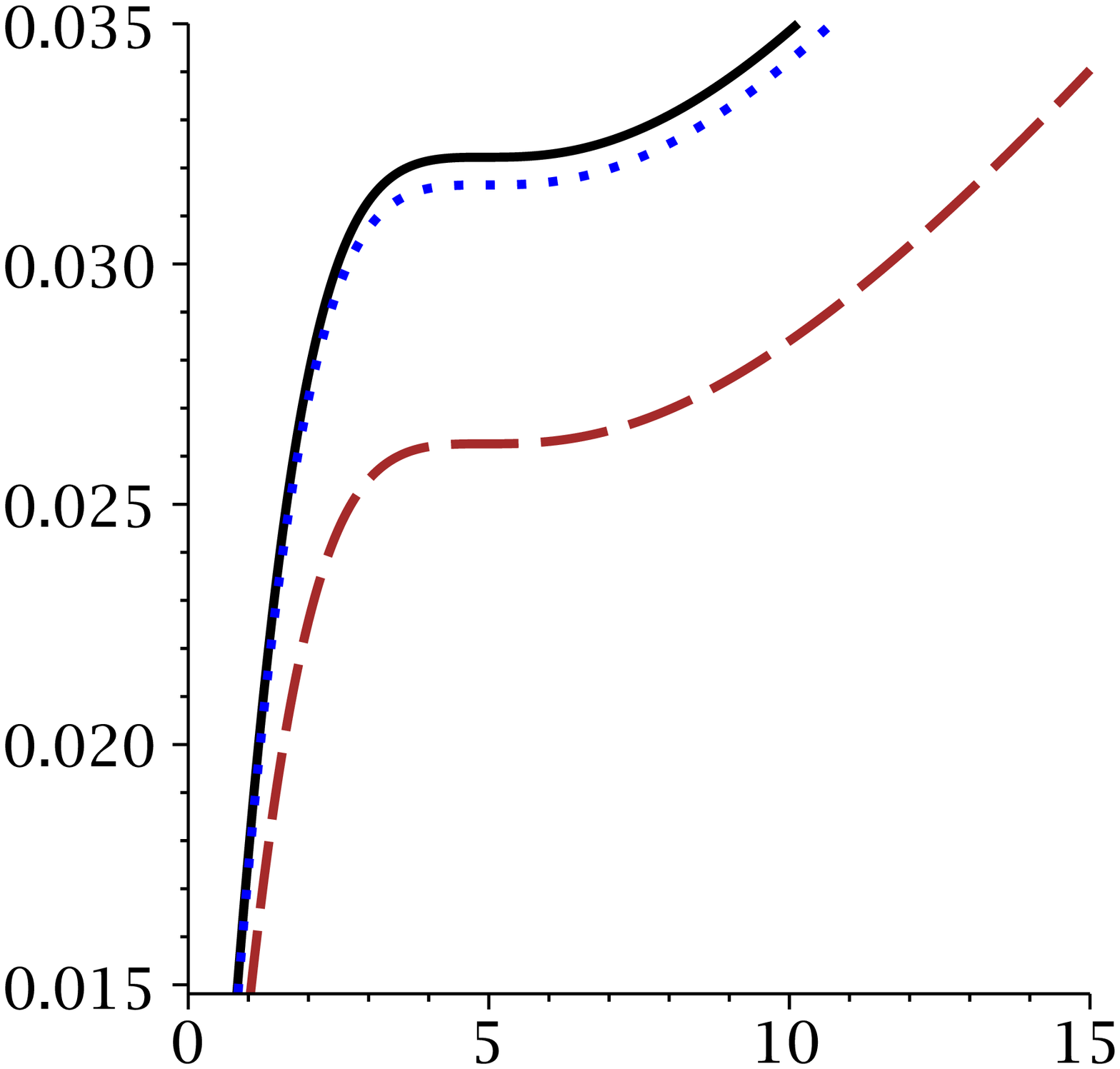} & \epsfxsize=5cm \epsffile{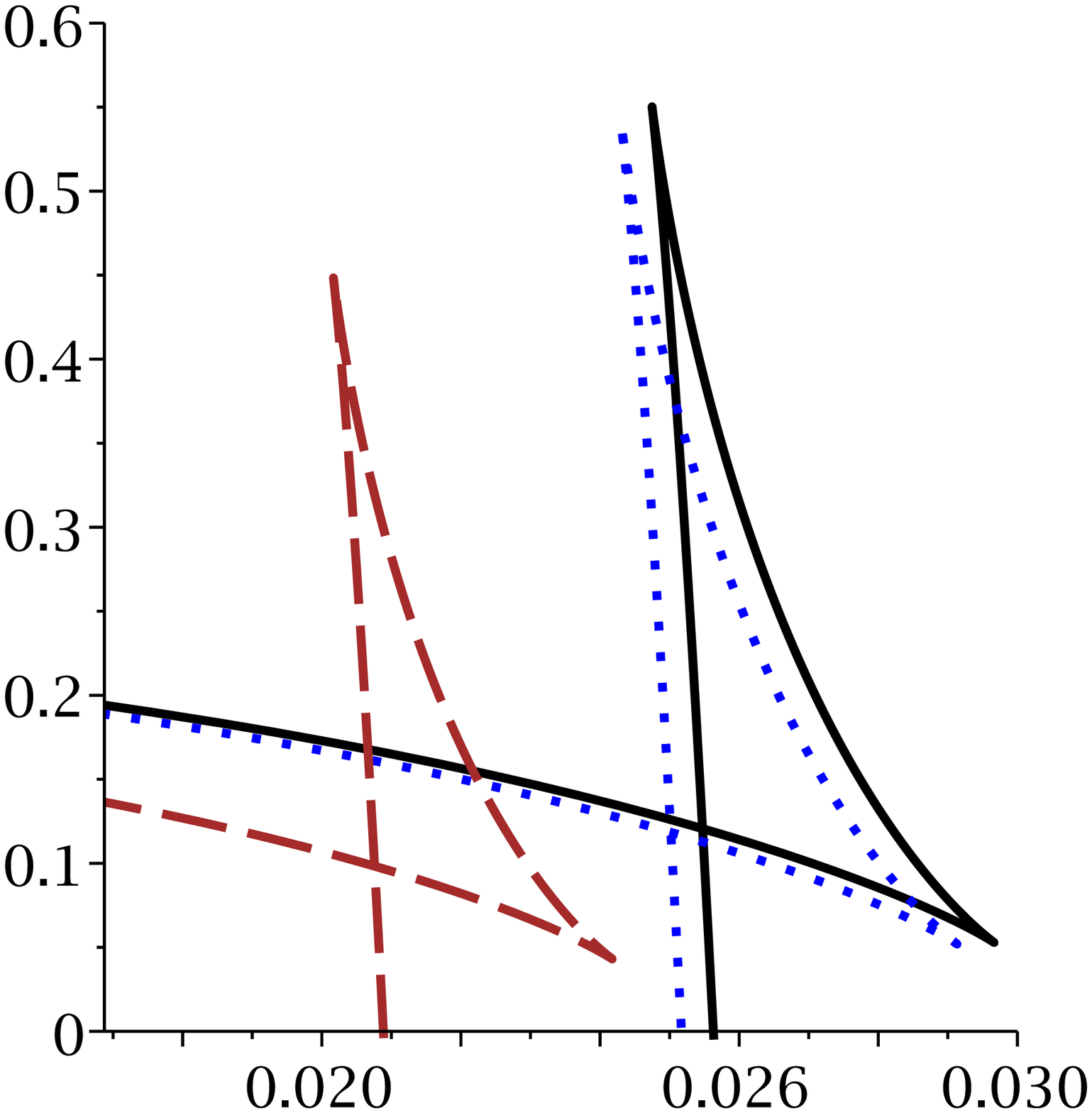}%
\end{array}
$%
\caption{ $P-r_{+}$ (left), $T-r_{+}$ (middle) and $G-T$ (right) diagrams
for $\protect\alpha ^{\prime }=5$, $q=0$ and $d=5$. \newline
$g\left( E/E_{p}\right) =1$, $f\left( E/E_{p}\right) =\frac{e^{\protect\beta %
E/E_{p}}-1}{\protect\beta E/E_{p}}$, $h_{i}(E)=0.9$, $E=1$, $E_{p}=5$, $%
\protect\beta=0.02$ (continuous line), $\protect\beta=0.2$ (dotted line) and
$\protect\beta=2$ (dashed line). \newline
$P-r_{+}$ diagram for $T=T_{c}$, $T-r_{+}$ diagram for $P=P_{c}$ and $G-T$
diagram for $P=0.5P_{c}$. }
\label{Fig2}
\end{figure}
%%%%%%%%%%%%%%%%%%%%%%%%%%%%%%%%%%%%%%%%%%%%%%%%%%%%%%%%%%%%%%%
%%%%%%%%%%%%%%%%%%%%%%%%%%%%%%%%%%%%%%%%%%%%%%%%%%%%%%%%%%%%%%%
\begin{figure}[tbp]
$%
\begin{array}{ccc}
\epsfxsize=5cm \epsffile{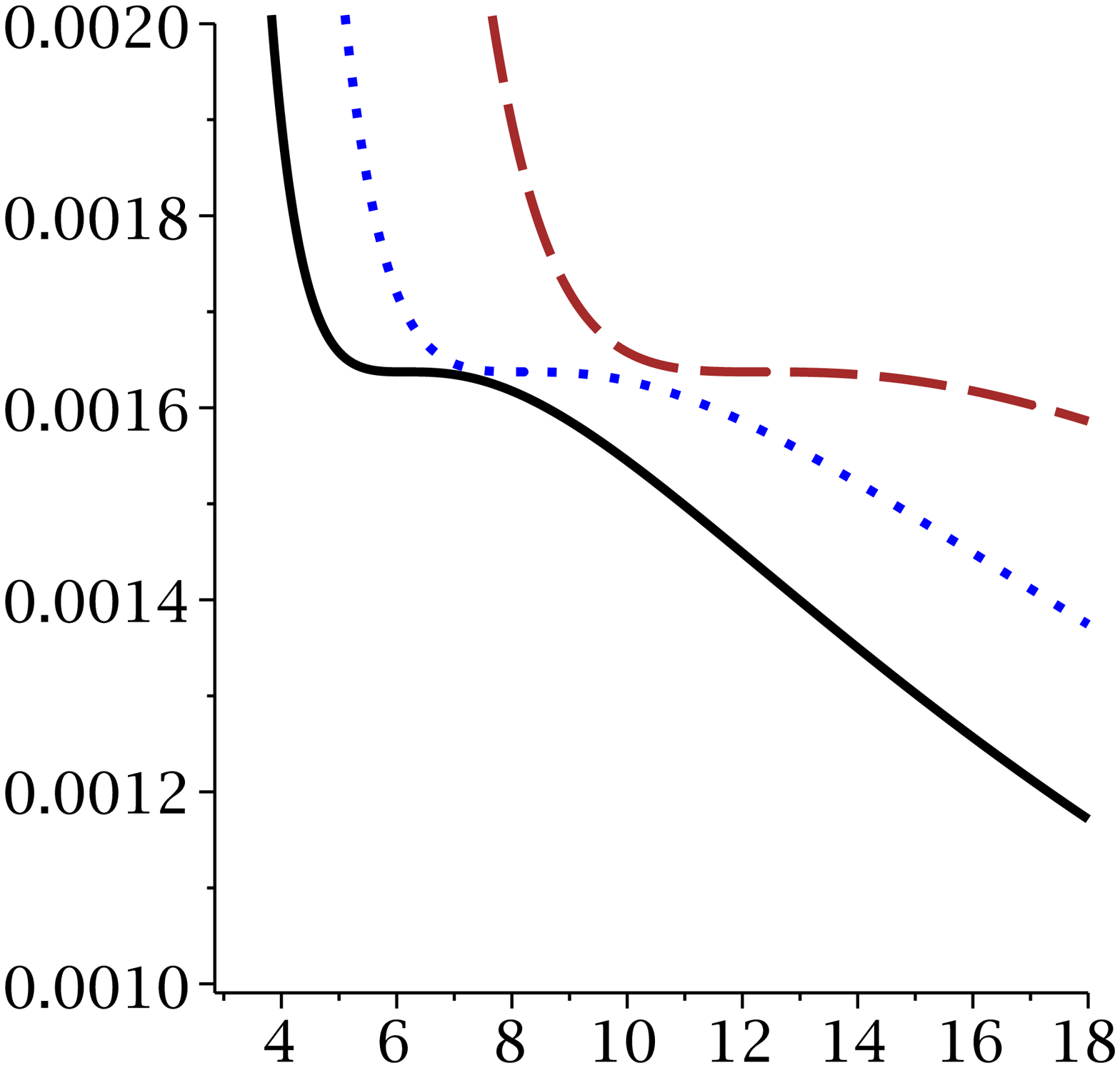} & \epsfxsize=5cm %
\epsffile{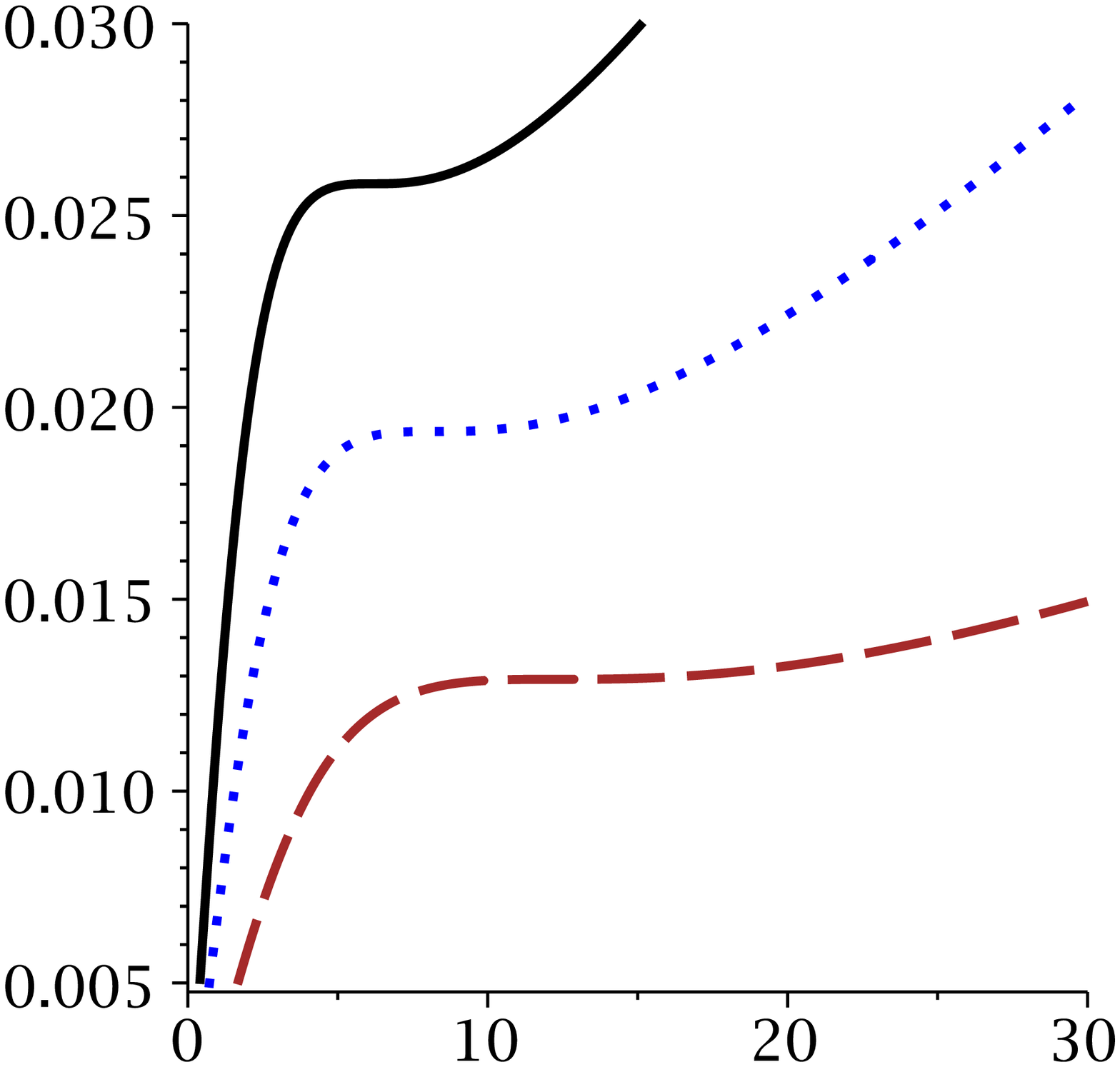} & \epsfxsize=5cm \epsffile{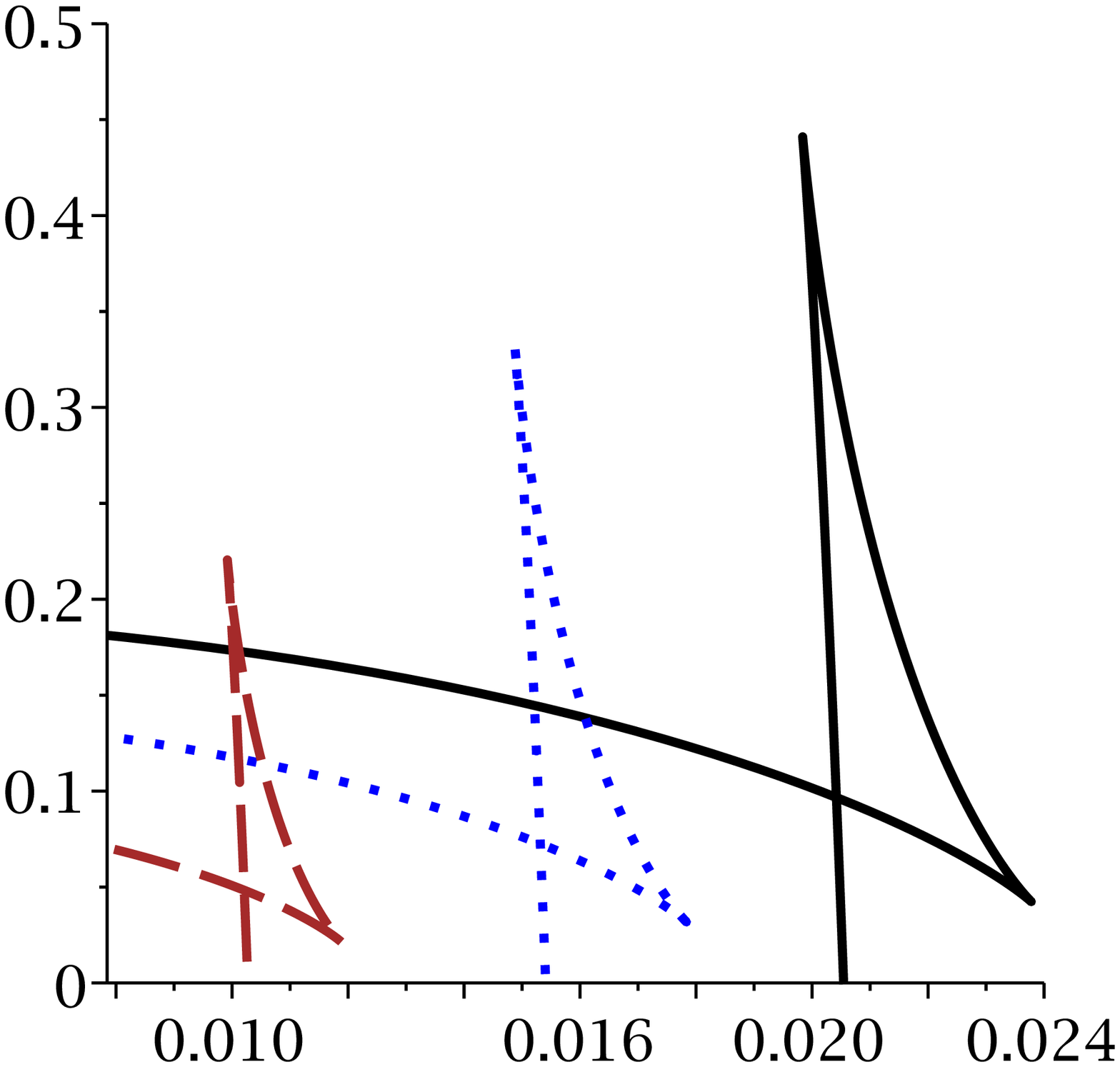}%
\end{array}
$%
\caption{ $P-r_{+}$ (left), $T-r_{+}$ (middle) and $G-T$ (right) diagrams
for $\protect\alpha ^{\prime }=5$, $q=0$ and $d=5$. \newline
$f\left( E/E_{p}\right) =g\left( E/E_{p}\right) =\frac{1}{1-\protect\lambda %
E/E_{p}}$, $h_{i}(E)=0.9$, $E=1$, $E_{p}=5$, $\protect\lambda=1$ (continuous
line), $\protect\lambda=2$ (dotted line) and $\protect\lambda=3$ (dashed
line). \newline
$P-r_{+}$ diagram for $T=T_{c}$, $T-r_{+}$ diagram for $P=P_{c}$ and $G-T$
diagram for $P=0.5P_{c}$. }
\label{Fig3}
\end{figure}
%%%%%%%%%%%%%%%%%%%%%%%%%%%%%%%%%%%%%%%%%%%%%%%%%%%%%%%%%%%%%%%
%%%%%%%%%%%%%%%%%%%%%%%%%%%coex%%%%%%%%%%%%%%%%%%%%%%%%%%%%%%%%%%%%
\begin{figure}[tbp]
$%
\begin{array}{ccc}
\epsfxsize=5cm \epsffile{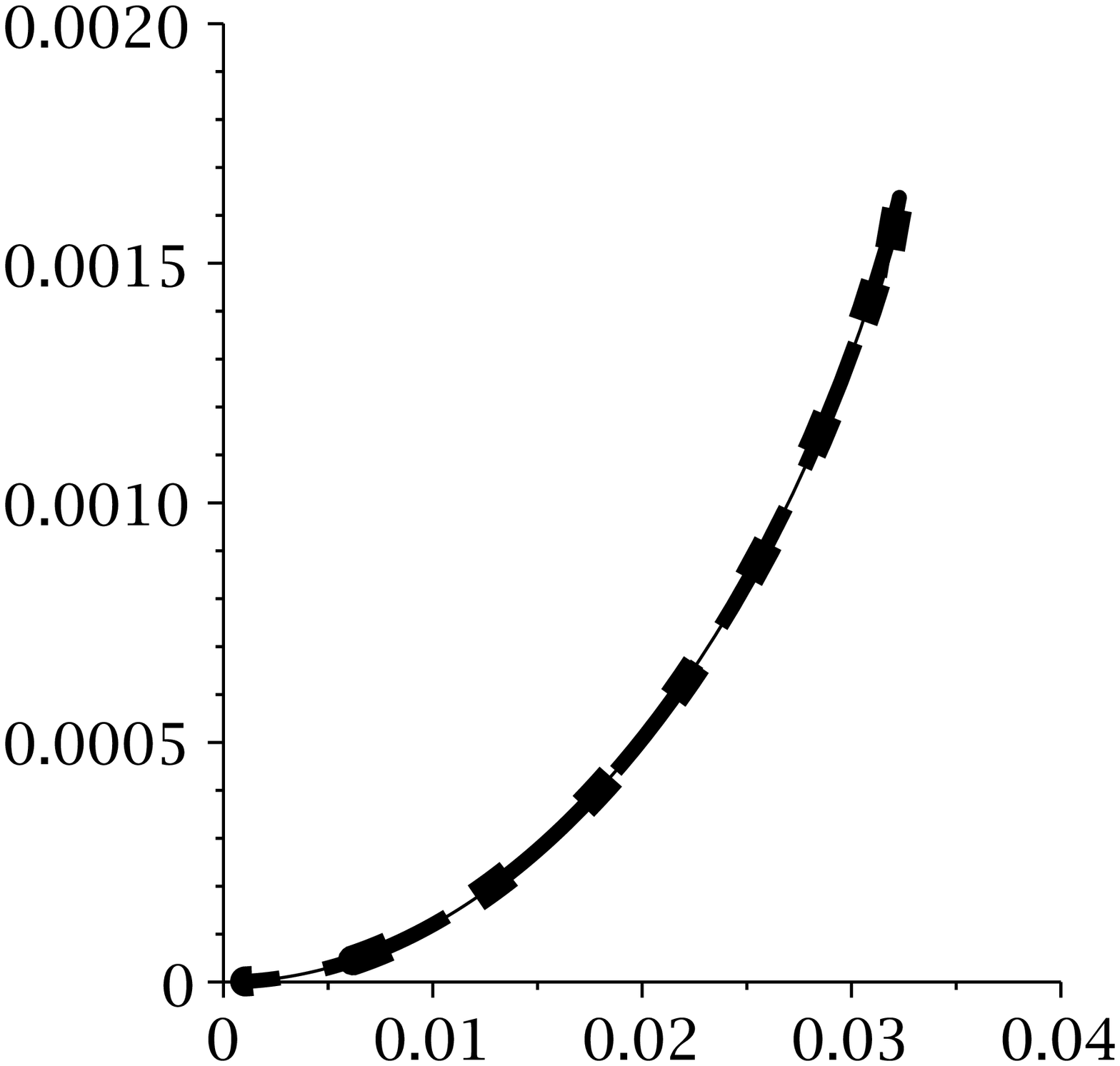} & \epsfxsize=5cm \epsffile{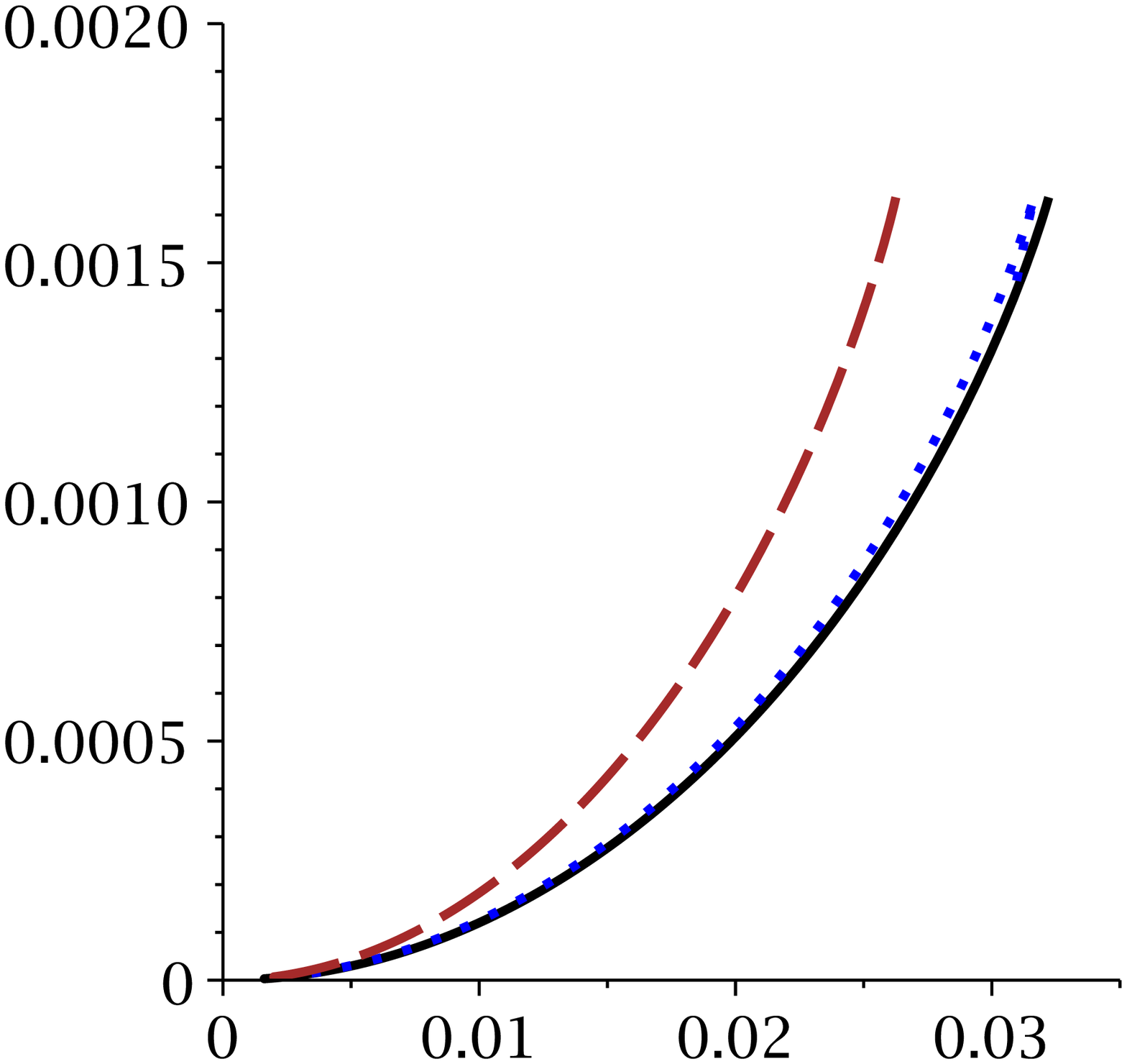}
& \epsfxsize=5cm \epsffile{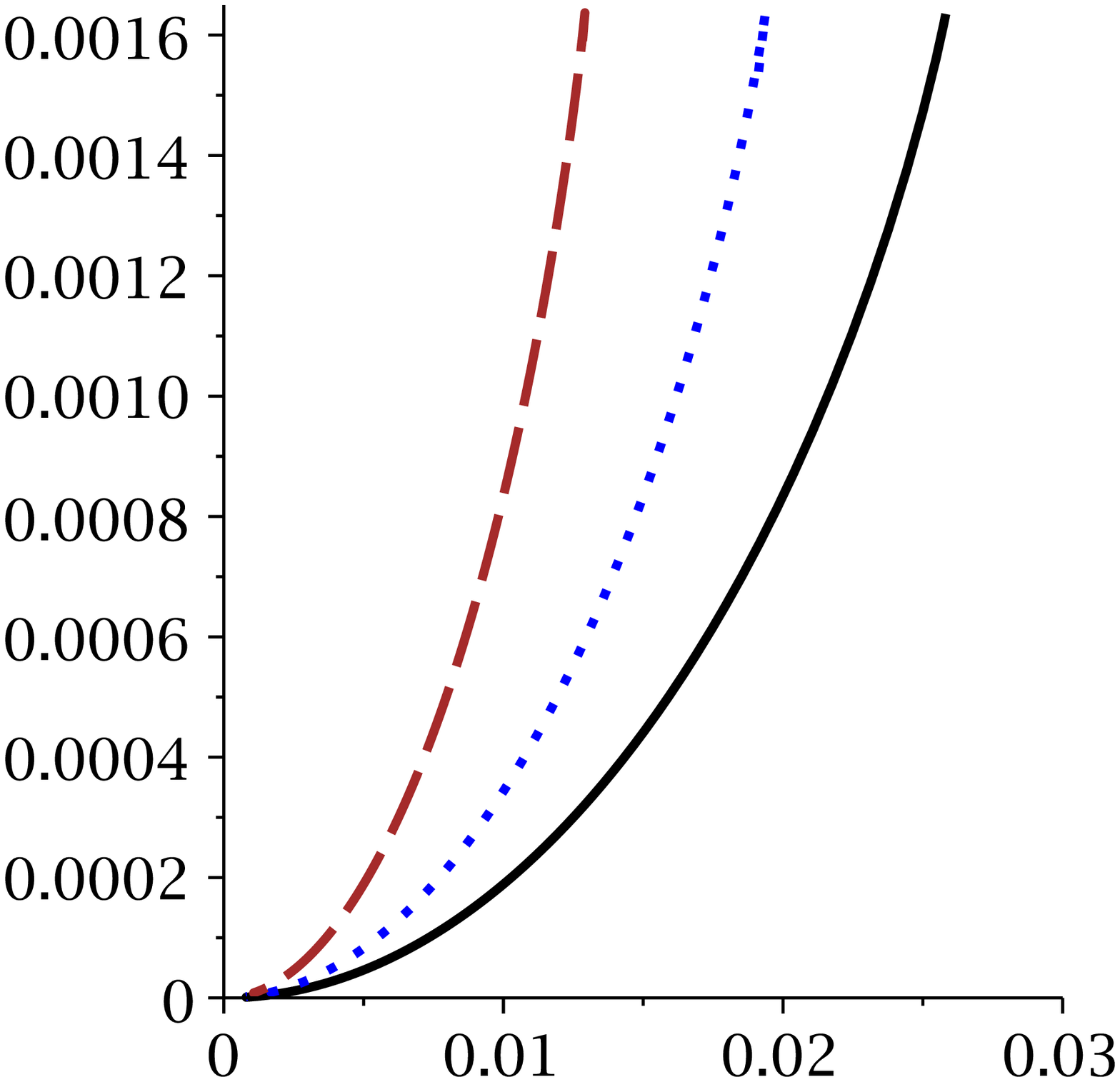}%
\end{array}
$%
\caption{ $P-T$ diagrams (left: $g\left(E/E_{p}\right)
=\protect\sqrt{1-\protect\eta \left( E/E_{p}\right) ^{n}}$,
$f\left( E/E_{p}\right) =1$), (middle: $g\left( E/E_{p}\right)
=1$, $f\left( E/E_{p}\right) =\frac{e^{\protect\beta
E/E_{p}}-1}{\protect\beta E/E_{p}}$) and (right: $f\left(
E/E_{p}\right) =g\left( E/E_{p}\right) =\frac{1}{1-\protect\lambda
E/E_{p}}$) for $\protect\alpha ^{\prime }=5$, $q=0$ , $d=5$,
$h_{i}(E)=0.9$, $E=1$ and $E_{p}=5$.
\newline For $g\left(E/E_{p}\right)
=\protect\sqrt{1-\protect\eta \left( E/E_{p}\right) ^{n}}$,
$f\left( E/E_{p}\right) =1$: $n=2$, $\protect\eta=1 $ (continuous
line), $\protect\eta=10$ (dotted line) and $\protect\eta=20$
(dashed line). \newline
 For $g\left( E/E_{p}\right) =1$, $f\left( E/E_{p}\right) =\frac{%
e^{\protect\beta E/E_{p}}-1}{\protect\beta E/E_{p}}$:
$\protect\beta=0.02$ (continuous line), $\protect\beta=0.2$
(dotted line) and $\protect\beta=2$ (dashed line). \newline
For $f\left( E/E_{p}\right) =g\left( E/E_{p}\right) =\frac{1}{1-%
\protect\lambda E/E_{p}}$: $\protect\lambda=1$ (continuous line), $\protect%
\lambda=2$ (dotted line) and $\protect\lambda=3$ (dashed line).}
\label{Coexq0}
\end{figure}
%%%%%%%%%%%%%%%%%%%%%%%%%%%%%%%%%%%%%%%%%%%%%%%%%%%%%%%%%%%%%%%
%%%%%%%%%%%%%%%%%%%%%%%%%%%%%%%%%%%%%%%%%%%%%%%%%%%%%%%%%%%%%%%
\begin{figure}[tbp]
$%
\begin{array}{ccc}
\epsfxsize=5cm \epsffile{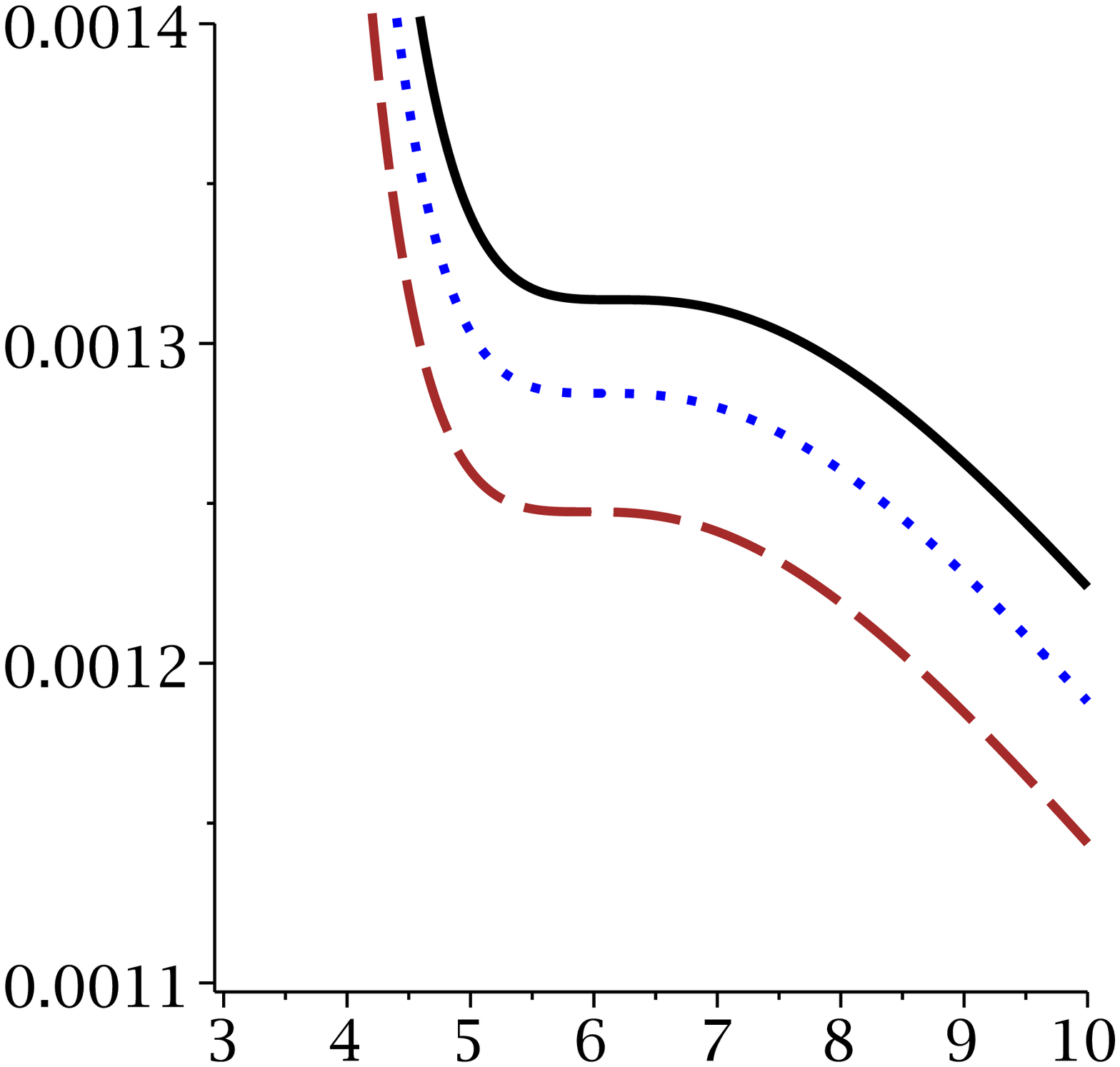} & \epsfxsize=5cm %
\epsffile{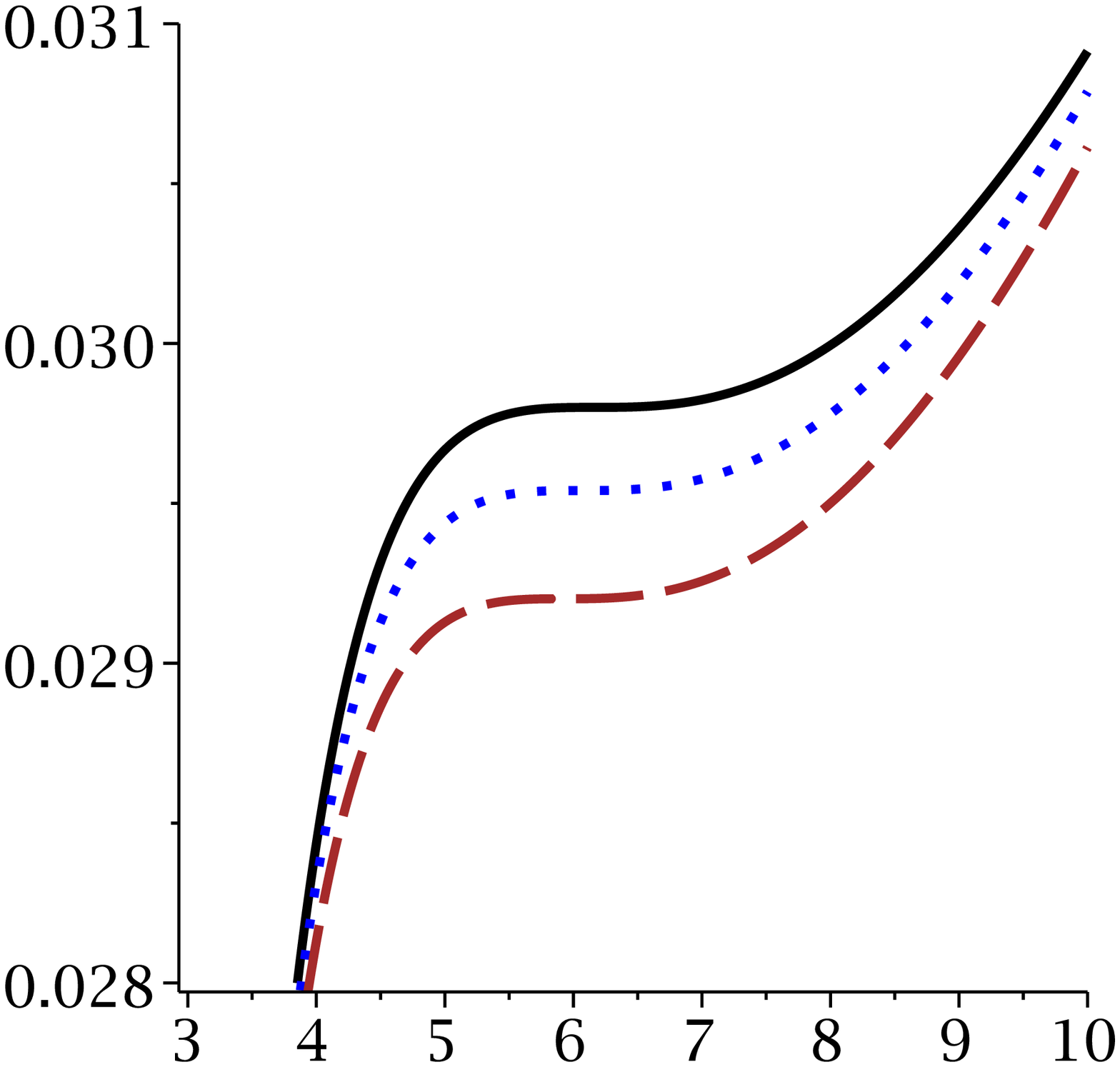} & \epsfxsize=5cm \epsffile{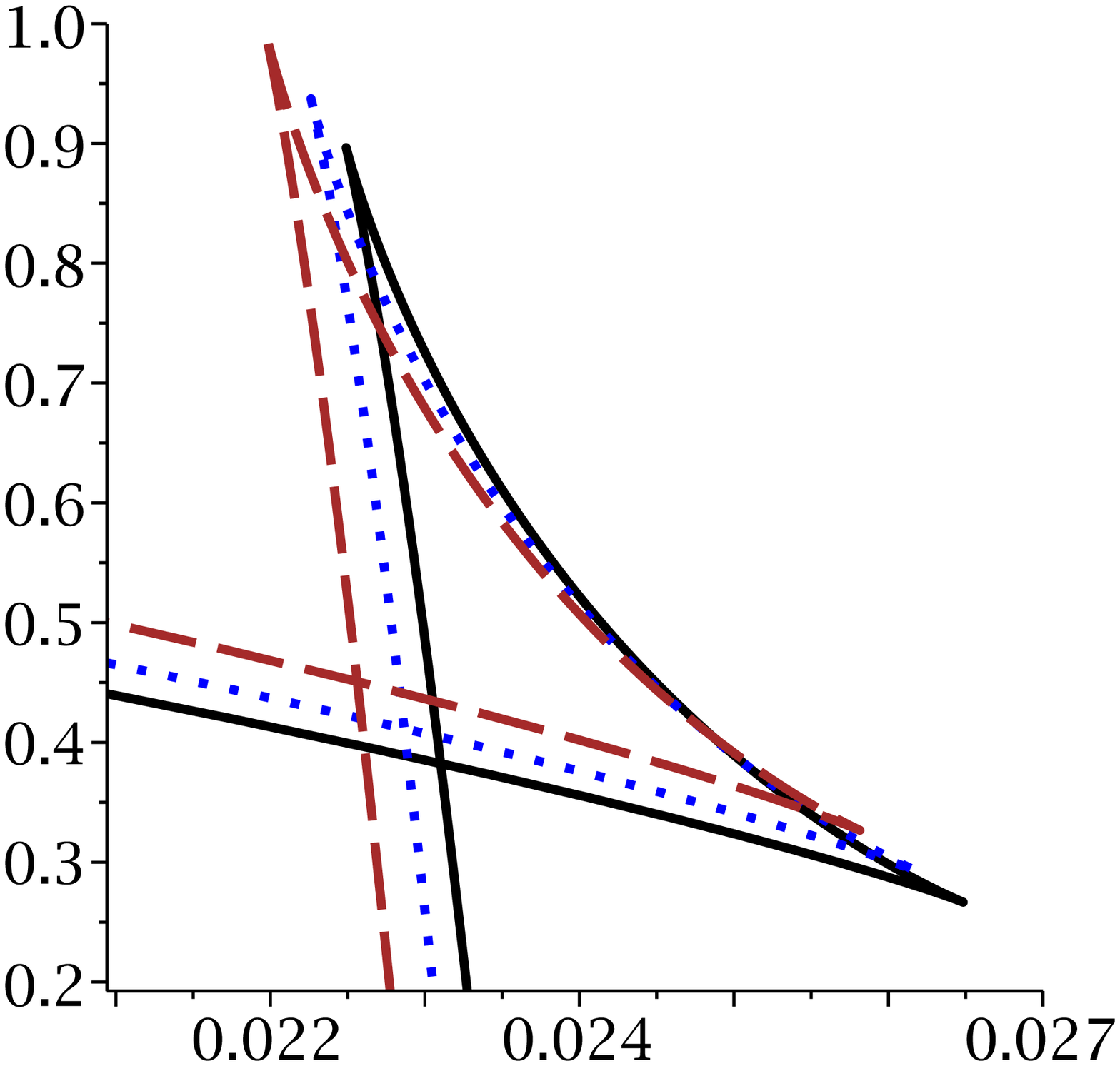}%
\end{array}
$%
\caption{ $P-r_{+}$ (left), $T-r_{+}$ (middle) and $G-T$ (right) diagrams
for $\protect\alpha ^{\prime }=5$, $q=6$ and $d=5$. \newline
$g\left(E/E_{p}\right) =\protect\sqrt{1-\protect\eta \left( E/E_{p}\right)
^{n}}$, $f\left( E/E_{p}\right) =1$, $h_{i}(E)=0.9$, $E=1$, $E_{p}=5$, $n=3$%
, $\protect\eta=1 $ (continuous line), $\protect\eta=10$ (dotted line) and $%
\protect\eta=20$ (dashed line). \newline
$P-r_{+}$ diagram for $T=T_{c}$, $T-r_{+}$ diagram for $P=P_{c}$ and $G-T$
diagram for $P=0.5P_{c}$. }
\label{Fig4}
\end{figure}
%%%%%%%%%%%%%%%%%%%%%%%%%%%%%%%%%%%%%%%%%%%%%%%%%%%%%%%%%%%%%%%
%%%%%%%%%%%%%%%%%%%%%%%%%%%%%%%%%%%%%%%%%%%%%%%%%%%%%%%%%%%%%%%
\begin{figure}[tbp]
$%
\begin{array}{ccc}
\epsfxsize=5cm \epsffile{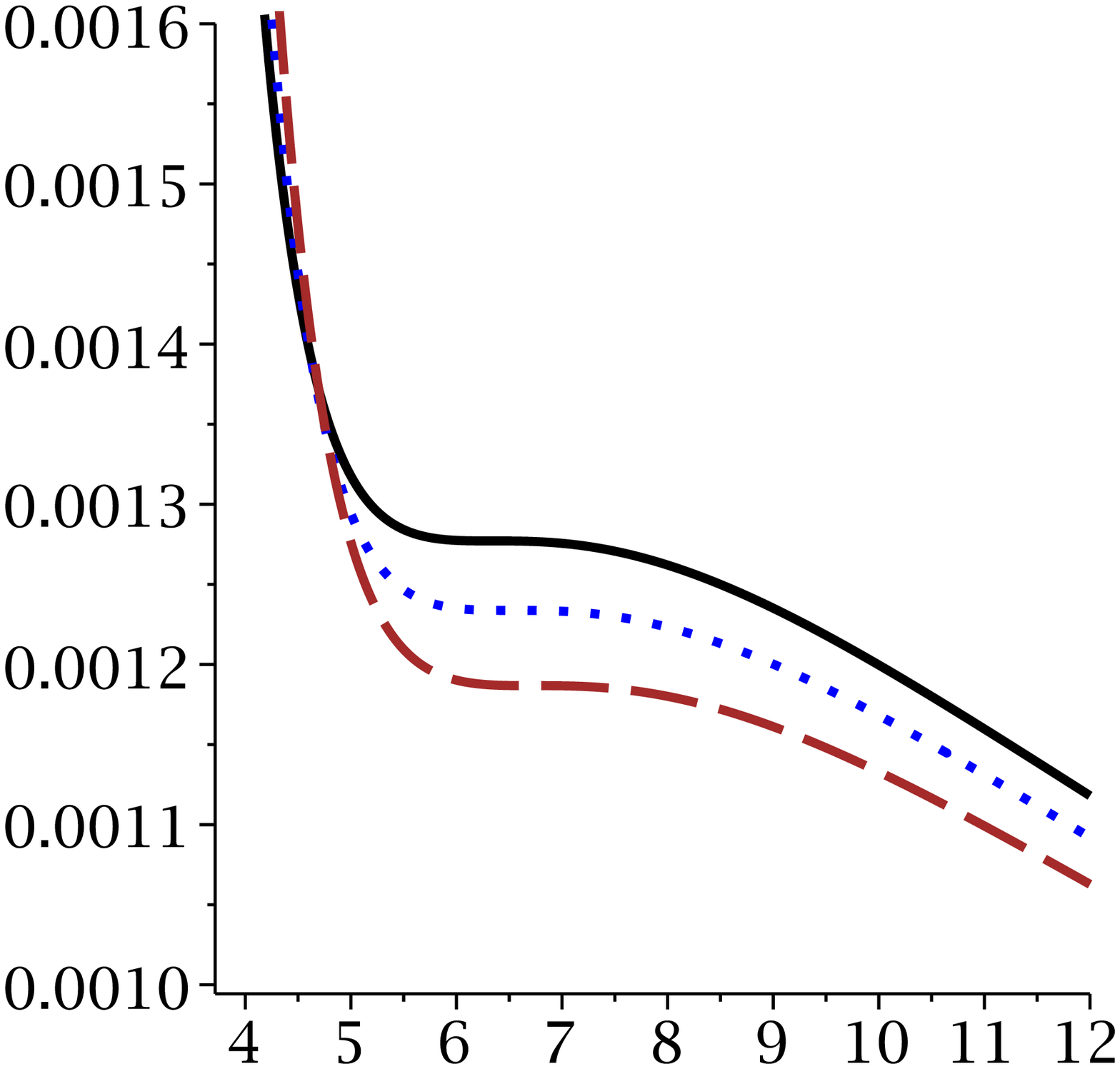} & \epsfxsize=5cm %
\epsffile{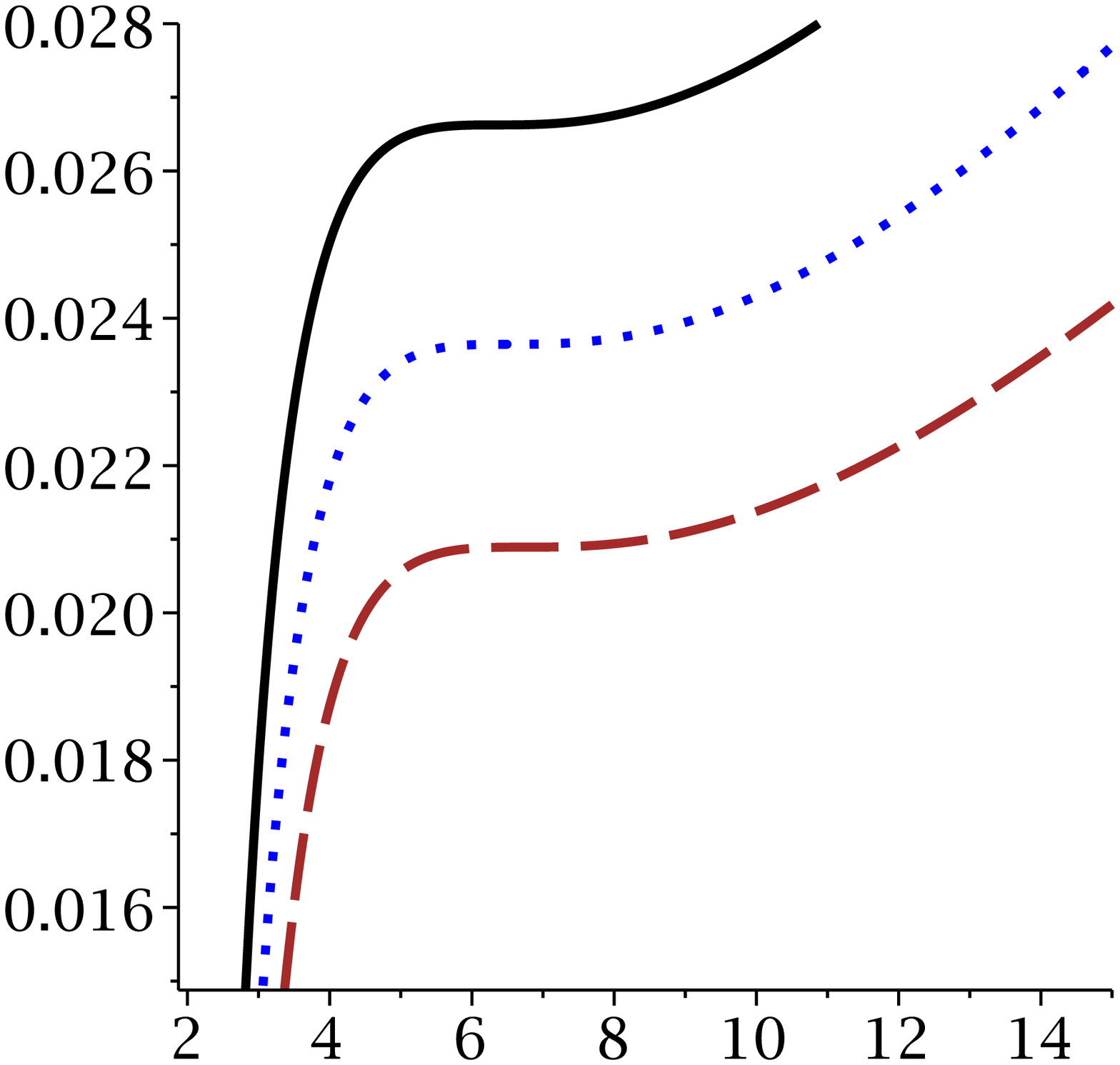} & \epsfxsize=5cm \epsffile{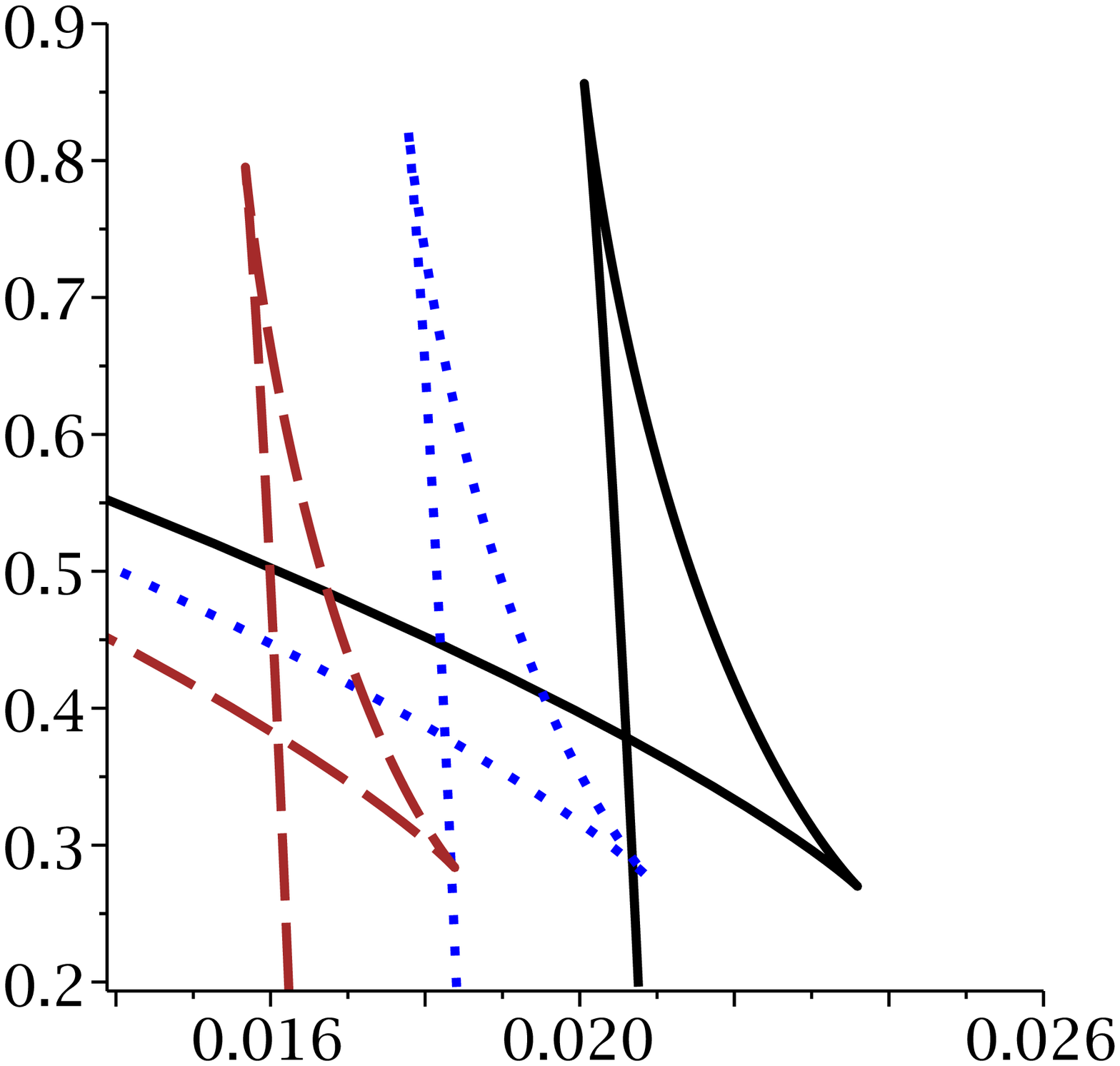}%
\end{array}
$%
\caption{ $P-r_{+}$ (left), $T-r_{+}$ (middle) and $G-T$ (right) diagrams
for $\protect\alpha ^{\prime }=5$, $q=6$ and $d=5$. \newline
$g\left( E/E_{p}\right) =1$, $f\left( E/E_{p}\right) =\frac{e^{\protect\beta %
E/E_{p}}-1}{\protect\beta E/E_{p}}$, $h_{i}(E)=0.9$, $E=1$, $E_{p}=5$, $%
\protect\beta=1$ (continuous line), $\protect\beta=2$ (dotted line) and $%
\protect\beta=3$ (dashed line). \newline
$P-r_{+}$ diagram for $T=T_{c}$, $T-r_{+}$ diagram for $P=P_{c}$ and $G-T$
diagram for $P=0.5P_{c}$. }
\label{Fig5}
\end{figure}
%%%%%%%%%%%%%%%%%%%%%%%%%%%%%%%%%%%%%%%%%%%%%%%%%%%%%%%%%%%%%%%
%%%%%%%%%%%%%%%%%%%%%%%%%%%%%%%%%%%%%%%%%%%%%%%%%%%%%%%%%%%%%%%
\begin{figure}[tbp]
$%
\begin{array}{ccc}
\epsfxsize=5cm \epsffile{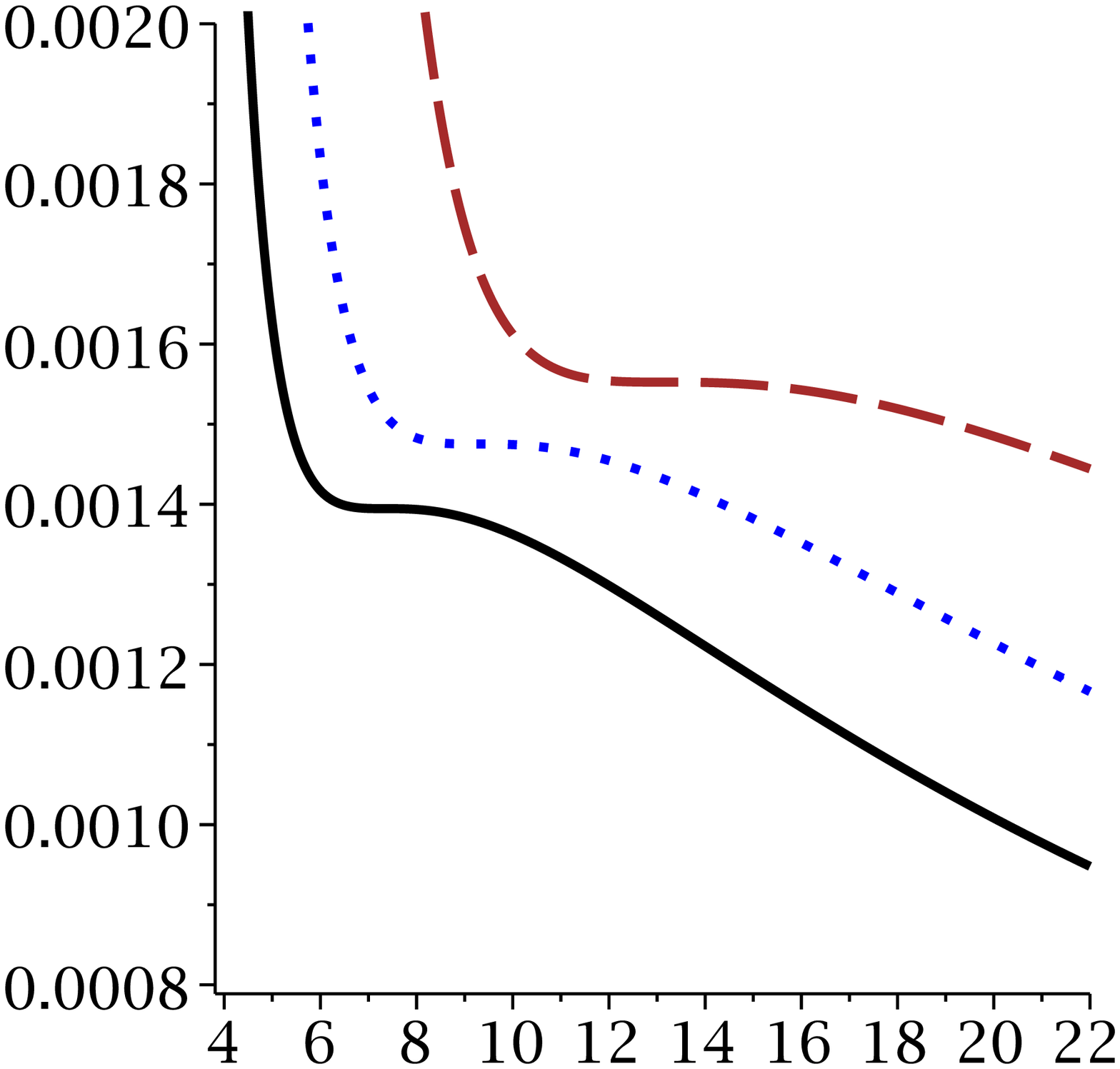} & \epsfxsize=5cm %
\epsffile{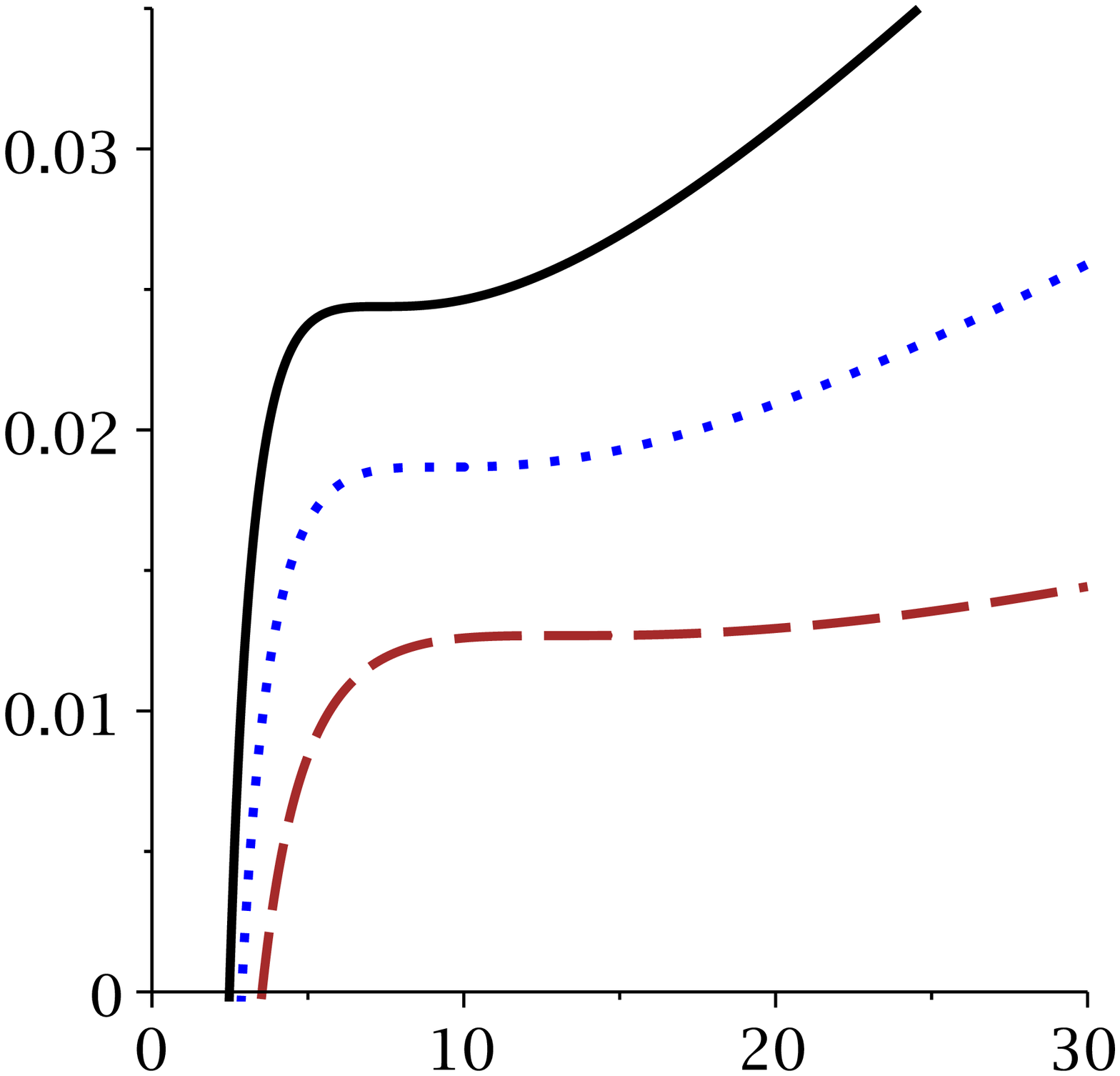} & \epsfxsize=5cm \epsffile{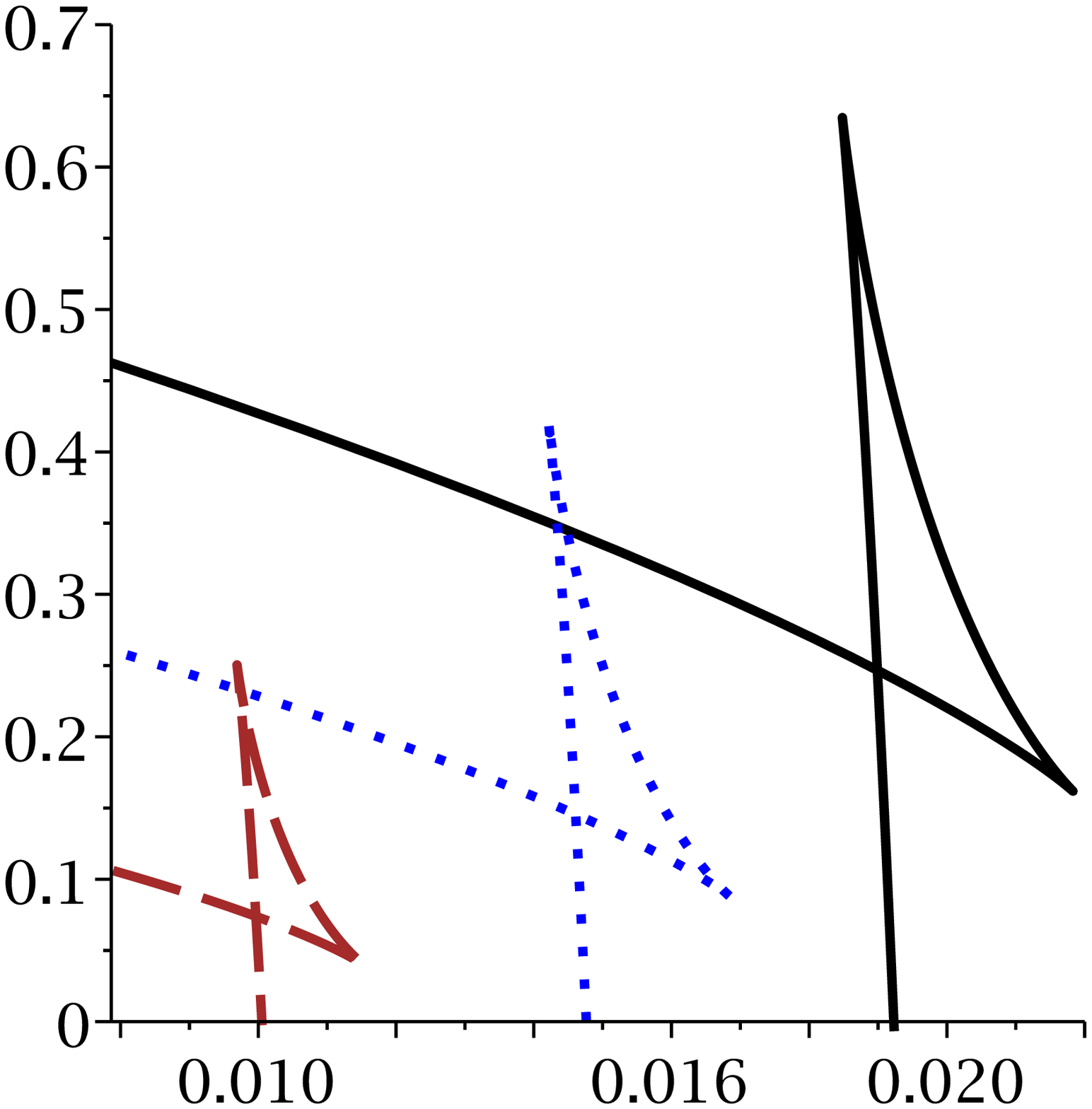}%
\end{array}
$%
\caption{ $P-r_{+}$ (left), $T-r_{+}$ (middle) and $G-T$ (right) diagrams
for $\protect\alpha ^{\prime }=5$, $q=6$ and $d=5$. \newline
$f\left( E/E_{p}\right) =g\left( E/E_{p}\right) =\frac{1}{1-\protect\lambda %
E/E_{p}}$, $h_{i}(E)=0.9$, $E=1$, $E_{p}=5$, $\protect\lambda=1$ (continuous
line), $\protect\lambda=2$ (dotted line) and $\protect\lambda=3$ (dashed
line). \newline
$P-r_{+}$ diagram for $T=T_{c}$, $T-r_{+}$ diagram for $P=P_{c}$ and $G-T$
diagram for $P=0.5P_{c}$. }
\label{Fig6}
\end{figure}
%%%%%%%%%%%%%%%%%%%%%%%%%%%%%%%%%%%%%%%%%%%%%%%%%%%%%%%%%%%%%%%
%%%%%%%%%%%%%%%%%%%%%%%%%%%%%%coex%%%%%%%%%%%%%%%%%%%%%%%%%%%%%%%%%
\begin{figure}[tbp]
$%
\begin{array}{ccc}
\epsfxsize=5cm \epsffile{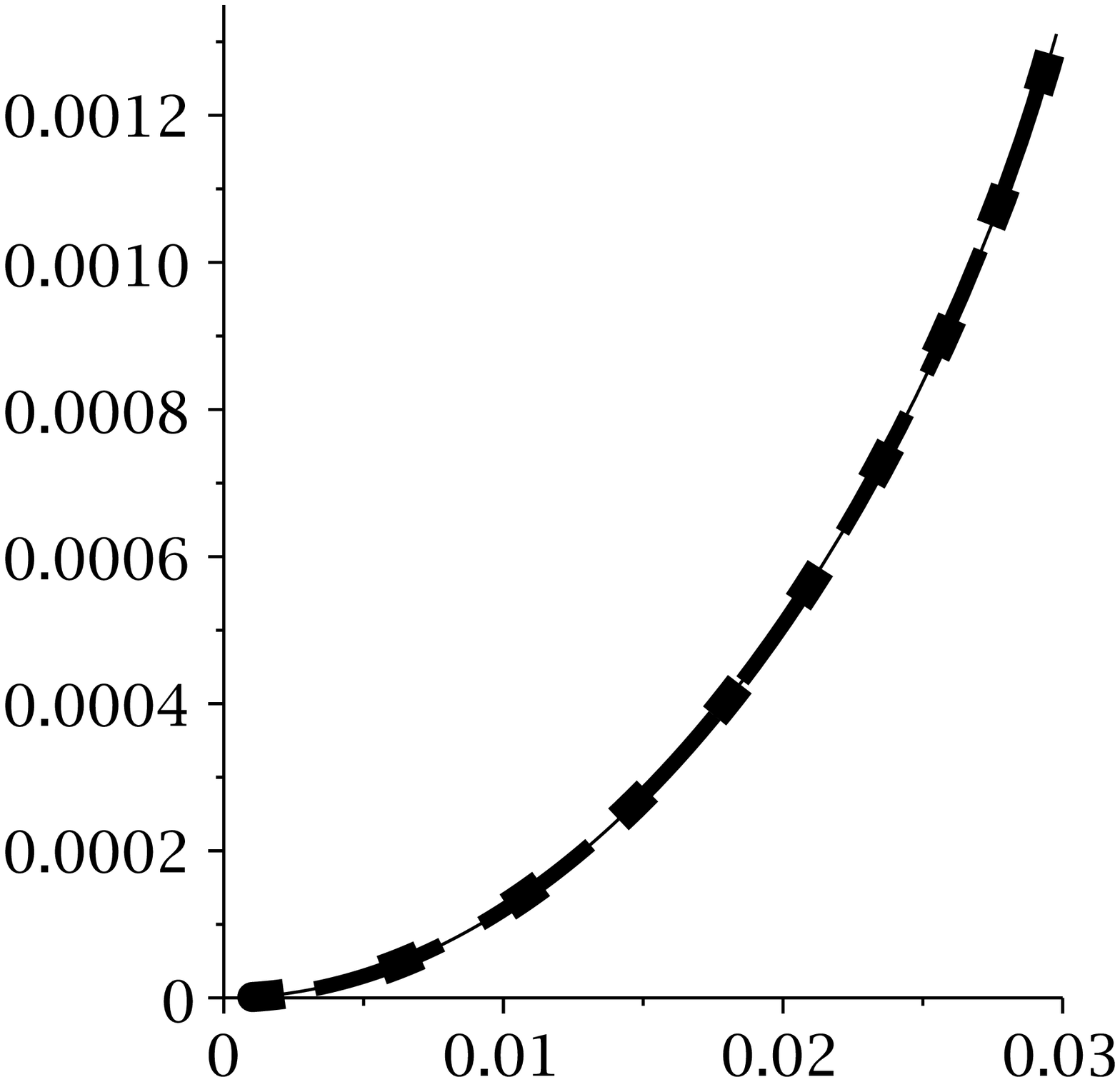} & \epsfxsize=5cm \epsffile{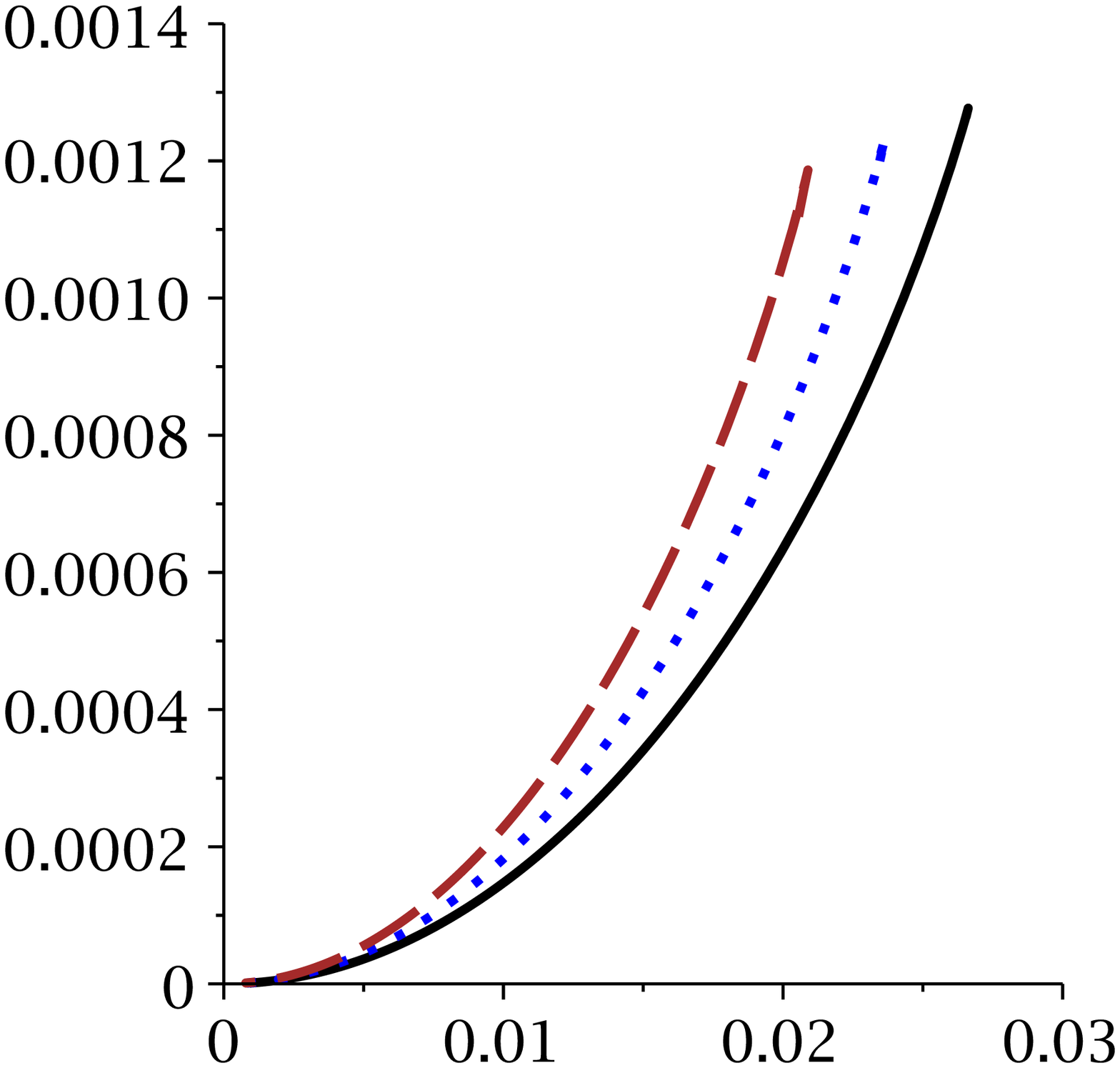}
& \epsfxsize=5cm \epsffile{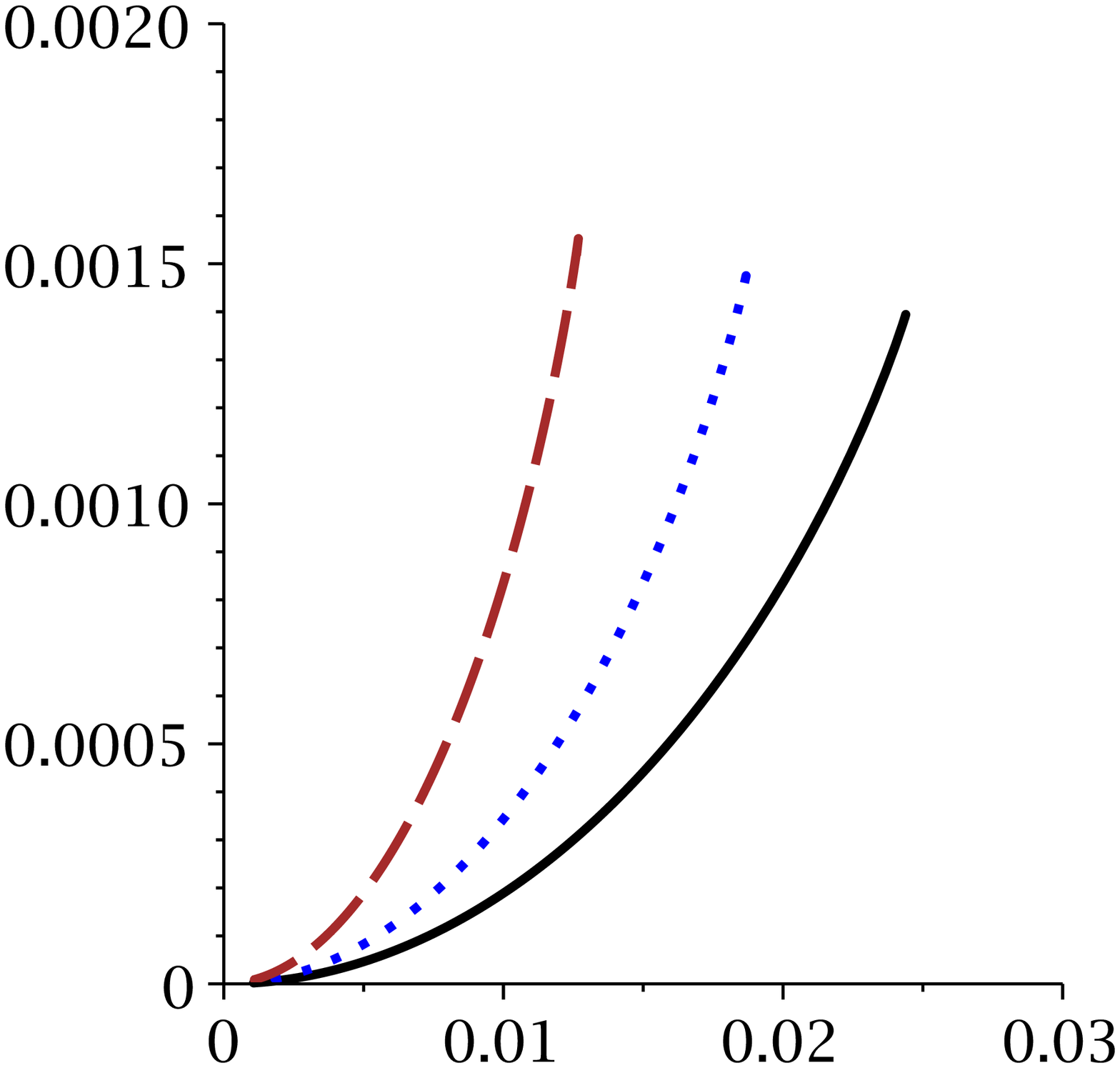}%
\end{array}
$%
\caption{ $P-T$ diagrams (left: $g\left(E/E_{p}\right)
=\protect\sqrt{1-\protect\eta \left(E/E_{p}\right) ^{n}}$,
$f\left( E/E_{p}\right) =1$), (middle: $g\left( E/E_{p}\right) =1$, $f\left( E/E_{p}\right) =\frac{%
e^{\protect\beta E/E_{p}}-1}{\protect\beta E/E_{p}}$) and (right:
$f\left( E/E_{p}\right) =g\left( E/E_{p}\right) =\frac{1}{1-%
\protect\lambda E/E_{p}}$) for $\protect\alpha ^{\prime }=5$,
$q=6$ , $d=5$, $h_{i}(E)=0.9$, $E=1$ and $E_{p}=5$. \newline For
$g\left(E/E_{p}\right) =\protect\sqrt{1-\protect\eta \left(
E/E_{p}\right) ^{n}}$, $f\left( E/E_{p}\right) =1$: $n=3$,
$\protect\eta=1 $ (continuous line), $\protect\eta=10$ (dotted
line) and $\protect\eta=20$ (dashed line). \newline
For $g\left( E/E_{p}\right) =1$, $f\left( E/E_{p}\right) =\frac{%
e^{\protect\beta E/E_{p}}-1}{\protect\beta E/E_{p}}$:
$\protect\beta=1$ (continuous line), $\protect\beta=2$ (dotted
line) and $\protect\beta=3$ (dashed line). \newline
For $f\left( E/E_{p}\right) =g\left( E/E_{p}\right) =\frac{1}{1-%
\protect\lambda E/E_{p}}$: $\protect\lambda=1$ (continuous line), $\protect%
\lambda=2$ (dotted line) and $\protect\lambda=3$ (dashed line).}
\label{Coexq}
\end{figure}
%%%%%%%%%%%%%%%%%%%%%%%%%%%%%%%%%%%%%%%%%%%%%%%%%%%%%%%%%%%%%%%

The obtained critical parameters indicate that in general, these
black holes with consideration of energy dependent constants,
enjoy second order phase transition in their phase space. Although
the presence of second order phase transition is restricted to
satisfy specific conditions, in general the second order phase
transition is a part of thermodynamical properties of these black
holes. Now, in order to elaborate the existence of second order
phase transition, we choose specific examples for free parameters
and energy functions to plot three set of diagrams which are
$T-r_{+}$, $P-r_{+}$ and $G-T$. The existences of subcritical
isobars for critical pressure in $T-r_{+}$ diagram, the region of
phase transition (reflection point) for critical temperature in
$P-r_{+}$ diagram and swallow-tail for pressures smaller than
critical pressure in $G-T$ diagram, indicate that a second order
phase transition is taking place for the obtained critical values.

Taking into account the rainbow functions, we regard three known models for
considering their effects. The first model is motivated from loop quantum
gravity and non-commutative geometry, in which rainbow functions are \cite%
{LQG}
\begin{equation}
f\left( E/E_{p}\right) =1,\text{ \ \ }g\left( E/E_{p}\right) =\sqrt{1-\eta
\left( E/E_{p}\right) ^{n}}.  \label{loop}
\end{equation}

The second model comes from the hard spectra of gamma-rays motivation with
the following form \cite{Amelino}
\begin{equation}
f\left( E/E_{p}\right) =\frac{e^{\beta E/E_{p}}-1}{\beta E/E_{p}},\text{ \ \
}g\left( E/E_{p}\right) =1.  \label{gamma}
\end{equation}%
Taking the constancy of the velocity of the light into account, one can find
following relations for the rainbow functions as third model
\begin{equation}
f\left( E/E_{p}\right) =g\left( E/E_{p}\right) =\frac{1}{1-\lambda E/E_{p}}.
\label{light}
\end{equation}

Now, using Eqs. (\ref{TNEW}), (\ref{PPNEW}) and (\ref{GNEW}) with
the obtained critical values (Eqs. (\ref{RCNEW}) - (\ref{PCNEW})),
one can plot the following $T-r_{+}$, $P-r_{+}$ and $G-T$ diagrams
for different models of
rainbow functions in special cases (Eqs. (\ref{loop}) - (\ref{light})) (Figs. \ref{Fig1} - %
\ref{Fig3} and \ref{Fig4} - \ref{Fig6}).

It is evident from plotted graphs that the obtained critical
values are representing the second order phase transition points
(due to the characteristic behaviors of the different phase
diagrams). In some classes of rainbow functions, for neutral case,
variations of the parameters of
rainbow functions have no effect (see right and left panels of Figs. \ref%
{Fig1} and \ref{Fig2}, respectively). This specific behavior was not
observed in the charged solutions.

By excluding right panel of Fig. \ref{Fig1}, in other cases, the critical
temperature, the size of swallow-tail and differences between energy of
different phases are decreasing functions of the parameters of rainbow
functions ($\eta $, $\beta $ and $\lambda $) (see right panels of Figs. \ref%
{Fig2}, \ref{Fig3} and \ref{Fig4} -- \ref{Fig6}). Meanwhile, by excluding
left panel of Fig. \ref{Fig2}, the critical pressure is a decreasing
function of $\eta $ and $\beta $ (see left panels of Figs. \ref{Fig1}, \ref%
{Fig4} and \ref{Fig5}) and an increasing function of $\lambda $ (see left
panels of Figs. \ref{Fig3} and \ref{Fig6}). As for subcritical isobar,
except for the neutral case of loop motivated rainbow functions (see middle
panel of Fig. \ref{Fig1}), its length is an increasing function of the
parameters of rainbow functions (see middle panels of Figs. \ref{Fig2}, \ref%
{Fig3} and \ref{Fig4} -- \ref{Fig6}).

In the absence of charge, presence of the rainbow functions provides
interesting effects. Taking a closer look at the Fig. \ref{Fig1} shows that
while we are varying rainbow function which results into modification of the
$P-r_{+}$ diagram, the critical temperature remains fixed (Fig. \ref{Fig1}
middle and right panels), and interestingly, the total behavior of the Gibbs
free energy versus temperature is not affected by this variation either.
Same behavior is observed in Fig. \ref{Fig2} left panel. This property
enables us to modify different critical values while a specific one of them
remains unchanged.

Using the fact that the Gibbs free energy, temperature, and pressure of the
system are constant during the phase transition, we have plotted the
coexistence line of the black holes (Figs. \ref{Coexq0} and \ref{Coexq}).
The small and large black holes have identical temperature and pressure
along the coexistence line, and the critical points are located at the end
of the coexistence line where above these points the phase transition does
not occur. Fig. \ref{Coexq0} (left panel) shows that the coexistence lines
for variation of $\eta $ are completely identical, because $g\left(
E/E_{p}\right) $ has no effect on the critical pressure and temperature. On
the other hand, Fig. \ref{Coexq} (left panel) indicates that variation of $%
\eta $ does not have significant effect on the coexistence lines. According
to properties of the coexistence lines, one can conclude that for these
black holes the reentrant of phase transition does not take place.

\section{Closing Remarks}

In this paper, the dependency of all constants on energy functions
was considered, and charged GB black holes in the presence of
gravity's rainbow were studied. Thermodynamical and geometrical
properties of these black holes were investigated and it was shown
that the power of the singularity, asymptotical behavior and
thermodynamical quantities of these black holes were modified due
to the dependency of constants on energy.

Next, using the concepts of extended phase space, van der Waals
like behavior and the second order phase transition of these black
holes were studied.

First of all, we found that in the presence of gravity's rainbow,
thermodynamical volume of the black holes is modified and it is
rainbow function dependent. In other words, the total behavior of
the volume is determined by rainbow functions as well as
dimensions.

Next, we found that, for neutral case, the obtained critical
radius and critical temperature were functions of rainbow
functions of the metric, whereas the critical pressure was
independent of them and was only dependent on energy variation of
the constants, which in return resulted in specific behaviors in
phase diagrams in special cases.

Interestingly, in the presence of the electric charge, a
limitation was found for having real positive critical horizon
radius. In other words, the presence of charge puts restriction on
values that different parameters can acquire. Contrary to the
neutral case, in the case of charged solutions, all critical
values (the critical temperature, the horizon radius and the
pressure) were dependent on rainbow functions of the metric. In
other words, the critical behavior of the system was modified due
to the presence of gravity's rainbow. It is worth mentioning that
the presence of rainbow functions was observed in ratio of
$\frac{P_{c}r_{c}}{T_{c}}$ for both charged and neutral cases,
which is a variation of a similar ratio in van der Waals system of
liquid/gas. In addition, the coexistence lines indicated that for
this type of black holes, the reentrant of phase transition does
not happen.

The specific behaviors and results of the paper motivate one to
analyze new interesting phenomenology for such black hole
thermodynamics. It will be worthwhile to study the effects of
non-linear electrodynamics on critical behavior of GB gravity's
rainbow. It will be interesting to generalize the obtained static
solutions to a case of dynamical ones and investigate the
cosmological consequences of such solutions. The behavior of the
Hawking radiation near the critical point is another subject of
interest.

\begin{acknowledgements}
The author would like to thank anonymous referees for suggesting
important improvements. We thank Shiraz University Research
Council. This work has been supported financially by the Research
Institute for Astronomy and Astrophysics of Maragha, Iran.
\end{acknowledgements}

\end{document}